\documentclass[%
 reprint,
superscriptaddress,
 amsmath,amssymb,
 aps,
 pra,
longbibliography
]{revtex4-2}

\usepackage{graphicx}
\usepackage{dcolumn}
\usepackage{bm}
\usepackage{booktabs}
\usepackage{hyperref}

\newcommand{\ket}[1]{\left\lvert #1 \right\rangle}
\newcommand{\abs}[1]{\left|#1\right|}

\begin{document}

\preprint{APS/123-QED}

\title{Quantum-enabled active matter at the atomic scale}

\author{Sabrina Burgardt}
\author{Julian Feß}
\author{Alexander Guthmann}
\author{Silvia Hiebel}
\affiliation{Department of Physics and State Research Center OPTIMAS, RPTU University Kaiserslautern-Landau,\\ Erwin-Schrödinger-Stra{\ss}e 46, 67663 Kaiserslautern, Germany}

\author{Aritra K. Mukhopadhyay}
\affiliation{Institute for Condensed Matter Physics, Technical University of Darmstadt, Hochschulstra{\ss}e 8, 64285 Darmstadt, Germany}

\author{\\ Sangyun Lee}
\affiliation{Institute of Physics, Johannes-Gutenberg University Mainz, Staudingerweg 9, 55128 Mainz, Germany}

\author{Michael te Vrugt}
\affiliation{Institute of Physics, Johannes-Gutenberg University Mainz, Staudingerweg 9, 55128 Mainz, Germany}

\author{Benno Liebchen}
\affiliation{Institute for Condensed Matter Physics, Technical University of Darmstadt, Hochschulstra{\ss}e 8, 64285 Darmstadt, Germany}

\author{Hartmut L\"owen}
\affiliation{Institute of Theoretical Physics II: Soft Matter, Heinrich-Heine University D\"usseldorf,\\ Universitätsstra\ss e 1, 40225 D\"usseldorf, Germany}

\author{Raphael Wittkowski}
\affiliation{Department of Physics, RWTH Aachen University and DWI -- Leibniz Institute for Interactive Materials,\\ Forckenbeckstra{\ss}e 50, 52074 Aachen, Germany}

\author{Artur Widera}
\email[Contact author; E-mail: ]{widera@rptu.de}
\affiliation{Department of Physics and State Research Center OPTIMAS, RPTU University Kaiserslautern-Landau,\\ Erwin-Schrödinger-Stra{\ss}e 46, 67663 Kaiserslautern, Germany}

\date{\today}

\begin{abstract}
Active matter comprises particles that extract energy from their local environment and convert it into motion. Although active particles have been miniaturized down to the nanoscale, realizing activity at the fundamentally smaller scale of individual atoms remains an open challenge, where quantum effects become increasingly relevant. Here, we experimentally demonstrate that individual $^{133}$Cs atoms confined in an optical dipole trap extract energy from an ultracold bath of $^{87}$Rb atoms via quantum-mechanical spin interactions and convert it into active motion. We quantitatively reproduce the resulting dynamics using a parameter-free active Langevin model derived from kinetic theory and support it with event-driven Monte Carlo collision simulations. The microscopic origin of activity is identified as quantum spin exchange, which transfers discrete internal spin energy into kinetic motion. Our work establishes a quantum-enabled route to active matter at the fundamental size limit of single atoms and opens perspectives for exploring the interplay of activity, quantum physics, and mesoscopic non-equilibrium thermodynamics.
\end{abstract}


\maketitle

\section{\label{sec1}Introduction}
The study of active matter -- systems driven out of equilibrium by a local influx and dissipation of energy -- has become one of the central research fields in statistical mechanics and soft matter over the past few decades \cite{MarchettiJRLPRS2013,BechingerdLLRVV2016}. 
Active matter science emerged from the study of biological systems, such as flocks of birds \cite{CavagnaG2014}, bacterial suspensions \cite{aranson2022bacterial}, or cytoskeleton biopolymers \cite{DoostmohammadiIYS2018}. 
Beyond the mere observation and understanding of active matter, an important and rapidly developing research direction is its artificial realization.
Artificial active matter spans over wide length scales, ranging from macroscopic implementations like artificial muscles \cite{shi2025ultrasound}, to microscopic objects like microrobots used for drug delivery \cite{landers2025clinically} or Janus colloids \cite{walther2008janus}.
Furthermore, despite significant challenges such as the massive enhancement of thermal fluctuations and surface effects when making active particles smaller, 
recent progress has made it possible to create active particles in the nanorealm \cite{Chen2025Nanomotors, CaleroSR2020}, opening up a whole new world of possibilities from medical applications and material science \cite{douglas2012logic,CiutiWKM2025} to nanoscale devices such as molecular robots and machines \cite{amir2014universal,WangEtAl2026}. 
All these systems are purely classical. 

Pushing miniaturization further to the atomic scale opens compelling new application potential. A dramatically increased density of active units, down to single atoms, would enable fundamentally new approaches to material design and engineering. In this regime, however, quantum effects can no longer be neglected and may even be harnessed as a resource. At the same time, such a reduction in scale challenges established activation principles, as thermal fluctuations dominate under typical conditions, requiring ultralow temperatures and densities. Consequently, there have been a number of theoretical proposals in recent years for realizations of quantum active matter. Examples include lattice models with flocking behavior \cite{AdachiTK2022,TakasanAK2024,rd46-hr3q}, mechanisms exploiting internal degrees of freedom \cite{YamagishiHO2024,PennerEtAl2025}, monitoring \cite{SteinervOE2026}, quantum particles driven along trajectories of classical active particles \cite{AntonovZLL2025,AntonovEtAl2026}, and spins with non-reciprocal interactions \cite{HanaiOT2025,NadolnyBB2025}. Moreover, there have been experiments with nonreciprocally coupled nanoparticles  \cite{RieserEtAl2022,ReisenbauerEtAl2024} and studies of objects swimming in superfluids \cite{KolmakovA2021,BakerRasooliEtAl2025}.

The combination of isolated, tightly controlled single atoms with a tailored dissipative coupling to a bath constitutes a paradigmatic open quantum system. Such systems can be realized and precisely controlled in ultracold atomic gases using ultrahigh vacuum, optical cooling and trapping techniques \cite{Massignan25}, together with single-atom, position- and spin-resolved detection \cite{Gross21}. A central challenge in this context is to identify microscopic mechanisms that enable activity under these conditions, fulfilling the defining criteria of active matter while remaining fully compatible with quantum mechanics.

\begin{figure*}[t!] 
\centering 
\includegraphics[width=\textwidth]{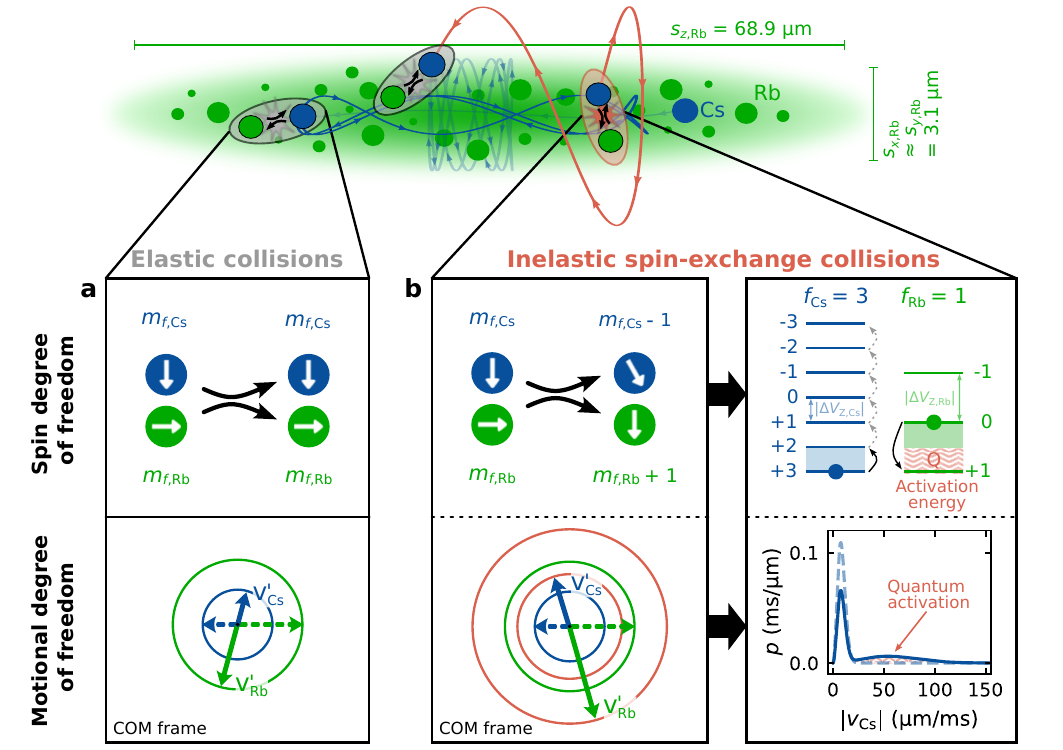} 
\caption{\textbf{Quantum mechanism breaking detailed balance}. A $^{133}\text{Cs}$ atom (blue) is immersed into a cigar-shaped, thermal bath of spin-polarized $^{87}\text{Rb}$ atoms (green) with extents $s_{x,\text{Rb}} \approx s_{y,\text{Rb}}$ and $s_{z,\text{Rb}}$ in the $x$-, $y$-, and $z$-direction, respectively. The in-trap motion of the Cs atoms (solid line) is influenced by elastic collisions (gray) and inelastic spin-exchange collisions (red) with the Rb atoms. The spin-exchange collisions cause an activation, i.e., impart kinetic energy to the Cs atoms (red trajectory), enabling them to leave the Rb bath and traverse the optical dipole trap before eventually re-entering the bath and thermalizing due to elastic collisions (blue trajectory). \textbf{a} Frequent elastic inter-species collisions do not alter the internal states of the colliding atom pair, and the kinetic energy of each atom is conserved in the center-of-mass (COM) frame. The velocities before and after the elastic collision in the COM frame are denoted with $\mathbf{v}$ and $\mathbf{v}^\prime$, respectively. 
\textbf{b} Inelastic inter-species spin-exchange collisions change the internal states $m_F$ of the atoms and release internal energy in quantized portions $Q$ to the relative kinetic energy, leading to a transient bimodal velocity distribution $p(\abs{v_{\text{Cs}}})$. The ratio of elastic and inelastic scattering rate ranges from $\Gamma_{\text{el}}/\Gamma_{\text{se}} \approx 25$ to 52.}  
\label{fig:experiment-sketch} 
\end{figure*}
In this work, we demonstrate that activity can indeed be realized at the level of individual atoms exploiting quantum effects. 
We show that individual $^{133}$Cs atoms are able to extract energy from a surrounding ultracold bath of thermal $^{87}$Rb atoms, and convert this energy into active motion (Fig.~\ref{fig:experiment-sketch}).
While such ultracold open quantum systems have been mainly used to experimentally study coherent system-bath interactions \cite{Grusdt2025, Baroni2024} or to probe the open system contribution on the quantum spin dynamics \cite{Wu2024}, here, by contrast, we exploit the spin-motion coupling to induce and control activation of the motional (i.e., kinetic) degree of freedom.

The activation mechanism relies on quantum spin-exchange in the $s$-wave regime, a process governed by total angular momentum and energy conservation \cite{Schmidt2018, Schmidt2019}. 
In each spin-exchange collision, internal (spin) energy from the surrounding Rb bath is converted into kinetic (motional) energy of the Cs atoms in quantized portions, resulting in an accelerating kick of the Cs atom in a random direction. 
This process therefore couples the spin and motional degrees of freedom microscopically in a quantized manner.
The dissipative character results in exo- and endothermal processes, which effectively break detailed balance at the level of individual inelastic collisions. 
This causes nonreciprocal transition rates and an effective energy flow from the bath that activates the Cs particle. 

This combination of discrete energy release and asymmetry of the interaction channels serves as a fundamentally new activation mechanism that has no counterpart in classical microswimmers, where typically a biased conversion of stored energy leads to directed kicks.
Additionally, this mechanism is microscopically fundamentally different from the macroscopically driven degree of freedom in classical superelastic or driven collision systems \cite{LeBlaySM2025}.
Importantly, the activation energy originates from the spin degree of freedom and is therefore decoupled from the motional (thermal) degree of freedom; the thermal energy of the surrounding gas does not constitute the fuel driving activation. 
These properties set the system apart from a mere heated impurity in a gas, but rather create a tunable nonequilibrium state.

We observe a strongly enhanced width of the Cs atoms’ density distribution upon activation that can be tuned via the activation energy and can by far exceed the width of the surrounding Rb cloud.
We contrast this behavior to the case of a passive Cs ensemble when the self-propulsion mechanism is blocked. 
The resulting long-time dynamics can be effectively described, first, by a coarse-grained Langevin model that 
creates an effective mapping to established active matter systems, underscoring the active nature of our atomic system. We derive this model from a fully microscopic kinetic description of the quantum transitions. 
Second, we develop an event-driven Monte-Carlo simulation describing the system on a microscopic level. 
In both models, we include crucial quantum effects of the atomic wave function in ultra-low energy scattering, which strongly boosts the activation far beyond the classical expectation.

Our single-atom-scale quantum-enabled active matter experiment opens the door towards a new generation of active particles, characterized by unprecedented smallness, unreached simplicity, and parameter tunability. The system offers a highly controllable platform for testing fundamental concepts of active matter, such as entropy production \cite{LoosK2020,NardiniEtAl2017}, time-reversal symmetry breaking, and activity-induced collective behaviors at ultrasmall scales and temperatures. Beyond that, our work provides a concrete pathway to realize active matter in the deep quantum regime in the near future and could serve as an avenue to enrich activity-induced phase transitions, such as flocking or motility-induced phase separation, with quantum effects.

\section{\label{sec2} Realizing active atom-sized particles}
We experimentally realize active atom-sized particles by immersing a small ensemble of $N_{\text{Cs}} = 68(2)$ neutral $^{133}\text{Cs}$ atoms into a bosonic, thermal $^{87}\text{Rb}$ cloud of $N_{\text{Rb}} = 7.7(1) \times 10^3$ atoms at $T_{\text{Rb}} = 463(10) \, \text{nK}$.
Both atomic species are co-trapped in a three-dimensional, anisotropic optical dipole trap (ODT). 
Details on the experimental sequence are given in App.~\ref{exp-setup}.
The activation of the diffusing Cs atoms is achieved via inelastic $s$-wave collisions with the Rb atoms.

Intuitively, the activated Cs atom statistically experiences a tunable acceleration in a random direction due to exothermal spin-exchange collisions as described below. 
These accelerating kicks persist until each Cs atom reaches its highest Zeeman state, after which it behaves as a passive diffusive particle.

\subsection*{Elastic and inelastic collisions}
The internal hyperfine ground state of Cs (Rb) acts as a quasi-spin with total angular momentum quantum number $f_{\text{Cs}} = 3$ ($f_{\text{Rb}} = 1$), and $m_{f,\text{Cs}}$ ($m_{f,\text{Rb}}$) is its projection onto the quantization axis given by the weak magnetic field with magnitude $B$.

Interactions between the two atomic species are determined by the central molecular interaction potential.
Its hyperfine interaction couples collisional channels which are defined by the total angular momentum $\mathbf{f}_{\text{tot}} = \mathbf{f}_{\text{Cs}} + \mathbf{f}_{\text{Rb}}$ and its projection $m_{\text{tot}}$ onto the quantization axis.
The interplay between the hyperfine interaction and the exchange interaction in the interaction Hamiltonian allows for two types of $s$-wave collisions \cite{Schmidt2018}:
First, elastic collisions, which preserve the internal state of the colliding atoms and redistribute kinetic energy between them; these lead to a thermalization of the motional degree of freedom between Rb and Cs.
Second, inelastic spin-exchange collisions, which result from the coupling of collisional channels of different spin states and can convert internal (spin) energy to kinetic (motional) energy.
These collisions lead to a quantized angular momentum transfer $\Delta m_f = \pm 1$ between the two atomic species
\begin{align}
    &\ket{1, m_{f,\text{Rb}}} \otimes \ket{3, m_{f,\text{Cs}}} \label{eq:spin-exchange}\\&\rightarrow  \ket{1, m_{f,\text{Rb}} - \Delta m_F} \otimes \ket{3, m_{f,\text{Cs}} + \Delta m_f} + Q,
    \notag
\end{align}
while maintaining the total angular momentum quantum number $f_{\text{tot}}$ and the projection $m_{\text{tot}} = m_{f,\text{Cs}} + m_{f,\text{Rb}}$.

\subsection*{Activation energy}
In addition to angular momentum, energy must also be conserved in every spin-exchange collision.
Internal energy $Q$ is released to (absorbed from) the collision energy $E_{\text{c}}$ during a spin-exchange collision with $\Delta m_f = -1$ ($\Delta m_f = +1$) while altering the internal state of the atoms, making the collision exothermal (endothermal).
The magnitude of the activation energy $Q$ is given by the total internal energy difference between the incoming and outgoing states
\begin{align}
    Q &= \Delta V_{\text{Z},\text{Rb}} + \Delta V_{\text{Z},\text{Cs}} \label{eq:Q} 
    \\&= \Delta m_f \left[ g_{f,\text{Cs}} - g_{f,\text{Rb}} \right] \mu_{\text{B}} B 
    \notag\\\notag& = -\Delta m_f k_{\text{B}} B \times 16.8 \, \mathrm{\text{\textmu} K \: G^{-1}}
\end{align}
where $\mu_{\text{B}}$ is the Bohr magneton and $k_{\text{B}}$ is the Boltzmann constant.
Throughout this paper, we express magnetic fields $B$ in gauss (G), where $1 \, \text{G} = 10^{-4} \, \text{T}$.
The energy $V_{\text{Z}} = m_f g_f \mu_{\text{B}} B$ is the Zeeman energy associated with the orientation of the quasi-spin $f$, and $g_{f,\text{Rb}} = -1/2 = 2 g_{f,\text{Cs}}$ denote the Land\'{e} factors of both atomic species, which differ by a factor of two.

\subsection*{Breaking detailed balance}
Exothermal spin-exchange collisions are energetically allowed for any magnetic field $B$.
These collisions represent a local energy input whose magnitude is set by $B$, resulting in a tunable strong acceleration of both collision partners in a random, but opposite, direction.
This is the mechanism that converts internal (spin) energy into kinetic (motional) energy of the colliding partners and drives the activation of the Cs atom. 
Subsequent elastic collisions of the collided atoms with the Rb cloud dissipate the acquired kinetic energy among the latter; this dissipation has a negligible effect on the Rb bath temperature due to the strong imbalance in atom numbers between the two atomic species.

The inverse process of endothermal spin-exchange collisions, however, is only possible if the collision energy $E_{\text{c}}$ of the colliding atom pair exceeds the energy $Q$.
The mean collision energy $\overline{E_{\text{c}}} = 3 k_{\text{B}} T_{\text{Rb}} / 2 \approx k_{\text{B}} \times 0.7 \, \mathrm{\text{\textmu} K}$ for a Rb bath temperature of $T_{\text{Rb}} = 463(10) \, \text{nK}$ is much smaller than the energy $\abs{Q} \approx k_{\text{B}} \times 17 \, \mathrm{\text{\textmu} K}$ to $k_{\text{B}} \times 51 \, \mathrm{\text{\textmu} K}$ for typical magnetic fields ranging from $1 \; \text{G}$ to $3 \; \text{G}$. 
As a result, endothermal spin-exchange collisions are practically absent, leading to a breakdown of local detailed balance and nonreciprocal transition rates that give rise to unidirectional spin exchange.
This breakdown is intrinsic to the quantum regime, where the discrete internal level structure and the ultralow collision energies render backward transitions effectively inaccessible.

A consequence of the broken detailed balance is that the system is brought to a markedly nonequilibrium state rather than only heating the impurity.
The finite number of internal Cs states restricts the maximum number of inelastic spin-exchange collisions per atom to six, i.e. activation stops once the uppermost Zeeman state has been reached.
This saturation distinguishes our quantum-enabled active particles from classical active matter, where propulsion is typically continuous or indefinite.

\subsection*{Quantum scattering}
On the microscopic scale, the local scattering rate of a single Cs atom at position $\mathbf{r}$ inside the Rb cloud is given by $\Gamma_{\text{tot}} = n_{\text{Rb}}(\mathbf{r}) \sigma_{\text{tot}}(E_{\text{c}}, B) v_{\text{c}}$ with local Rb density $n_{\text{Rb}}(\mathbf{r})$ and collision energy $E_{\text{c}} = \mu v_{\text{c}}^2/2$, where $v_{\text{c}}$ denotes the magnitude of the relative velocity. 
The reduced mass of the system is given by $\mu = m_{\text{Rb}} m_{\text{Cs}} / (m_{\text{Rb}} + m_{\text{Cs}})$, where $m_{\text{Rb}} = 86.9 \, \text{u}$ and $m_{\text{Cs}} = 132.9 \, \text{u}$ denote the mass of a Rb and a Cs atom, respectively.
The state-dependent scattering cross section $\sigma_{\text{tot}}(E_\text{c}, B) = \sigma_{\text{el}}(E_\text{c}, B) + \sigma_{\text{se}}(E_\text{c}, B)$ is the sum of the cross sections of all possible collision processes.

Different from classical active matter, all scattering cross sections crucially must be computed quantum mechanically for the matter wave function of the colliding atoms in the molecular interaction potential for the respective internal states. 
Such controlled-collision approaches have been used to probe quantum scattering of ultracold atoms at well-defined collision energies, including shape and Feshbach resonances in optical atom colliders \cite{Thomas2017QuantumScattering}.
Here, they are taken from coupled-channel scattering calculations that take into account the internal atomic structure of both species and the molecular interaction potential.
Importantly, the quantum nature of the system allows us to experimentally tune the inter-species interactions via $\sigma_{\text{tot}}(E_\text{c}, B)$ by varying the temperature $T_{\text{Rb}}$ (which sets the range of possible collision energies $E_{\text{c}}$) and the magnetic field $B$. 
Such controlled energy- and field-dependence of ultracold scattering has been exploited, e.g., to probe above-threshold Feshbach physics \cite{Horvath2017AboveThresholdFeshbach}.
Both the elastic and inelastic scattering cross sections decrease with increasing collision energy; hence, an activated particle with a strongly increased kinetic energy exhibits a reduced scattering cross section owing to its quantum wave function.
Thus, high-energy collisions are less probable, so that the collisional dynamics is much reduced, which effectively boosts the activation.

Moreover, the elastic and inelastic scattering cross sections differ in their magnetic field dependence: the elastic scattering cross section varies only very weakly with the magnetic field $B$, whereas the inelastic scattering cross section decreases with $B$. 
For typical experimental conditions, the ratio of elastic to inelastic scattering rate ranges from $\Gamma_{\text{el}}/\Gamma_{\text{se}} \approx 25$ to 52, indicating a strong thermalization of the Cs atoms between two spin-exchange collisions.
The precise knowledge of the $s$-wave scattering cross sections enables a direct comparison between our experimental data and numerical event-driven Monte-Carlo collision simulations, without any free parameters.
Further details on the scattering properties and these simulations are given in Apps.~\ref{scatteringproperties} and \ref{montecarlo}, respectively.

\subsection*{Emergence of a classical Markovian system}
The hierarchy of length scales summarized in Tab.~\ref{tab:scales} justifies treating the system as an underdamped, active atomic particle coupled to a Markovian bath.
\begin{table}
    \caption{\textbf{Typical experimental length scales.} The temperature $T_{\text{Rb}} = 463(10) \, \text{nK}$ of the Rb cloud is about a factor of two larger than the critical temperature of Bose-Einstein condensation. The mean-free path of the two atomic species has been calculated using the $s$-wave scattering cross sections $\sigma_{\text{el},\text{Cs}} \approx 2.45 \times 10^{-10} \, \text{cm}^2 \approx 34 \times \sigma_{\text{el},\text{Rb}}$ \cite{Kempen2002}.}
    \centering
    \begin{tabular}{ l l c } 
     \hline
    Cloud extent & $s_{x,\text{Rb}}$ & $3.1 \, \mathrm{\text{\textmu} m}$\\
         & $s_{y,\text{Rb}}$ & $3.1 \, \mathrm{\text{\textmu} m}$\\ 
     & $s_{z,\text{Rb}}$ & 68.9  $\mathrm{\text{\textmu} m}$\\
     Density & $\overline{n_{\text{Rb}}}$ & $0.28(2) \times 10^{13}$ $\text{cm}^{-3}$ \\
      & $n_{\text{Rb},\text{peak}}$ & $0.56(2) \times 10^{13} \, \text{cm}^{-3}$\\
     Inter-particle spacing & $\overline{n_{\text{Rb}}}^{-1/3}$ & $0.70 \, \mathrm{\text{\textmu} m}$\\
     & $n_{\text{Rb},\text{peak}}^{-1/3}$ & $0.56 \, \mathrm{\text{\textmu} m}$ \\
     de Broglie wavelength & $\lambda_{\text{th},\text{Rb}}$ & $0.27 \, \mathrm{\text{\textmu} m}$\\
      & $\lambda_{\text{th},\text{Cs}}$ & $0.22 \, \mathrm{\text{\textmu} m}$\\
      Mean-free path & $\Lambda_{\text{Rb}}$ & $340 \, \mathrm{\text{\textmu} m}$\\
      & $\Lambda_{\text{Cs}}$ & $10 \, \mathrm{\text{\textmu} m}$\\
     \hline
    \end{tabular}
    \label{tab:scales}
\end{table}
The relation between inter-particle spacing $\overline{n_{\text{Rb}}}^{-1/3}$ and the thermal de Broglie wavelength $\lambda_{\text{th},\text{Rb}} = h/\sqrt{2 \pi m_{\text{Rb}} k_{\text{B}} T_{\text{Rb}}}$ indicates that the Rb cloud can be described classically.
However, quantum effects could be enhanced by further cooling to increase the thermal de Broglie wavelength of Rb.
Furthermore, the mean-free path of Cs atoms, $\Lambda_{\text{Cs}} = 1/(\sqrt{2} \, \overline{n_{\text{Rb}}} \, \sigma_{\text{el},\text{Cs}})$, is significantly smaller than that of Rb atoms, yet remains on the order of the extents of the Rb cloud along all three spatial directions. 
This implies that Cs atoms probe a substantial fraction of the Rb cloud between collisions, while repeated collisions with the same Rb atom are highly unlikely. 
Consequently, the Rb cloud can be treated as a Markovian bath without memory.
Finally, the comparison of the Cs thermal de Broglie wavelength $\lambda_{\text{th},\text{Cs}}$ with the inter-particle spacing at peak density $n_{\text{Rb},\text{peak}}^{-1/3}$ shows that a classical description is also appropriate for the motion of the Cs atoms.

\section{Modeling active atom-sized particles}
To support our qualitative arguments that the Cs atoms are active particles, in App.~\ref{kinetictheory} we systematically derive a coarse-grained model for the dynamics of the Cs atoms. Remarkably, this model closely resembles established models from the active matter literature \cite{Fiasconaro_JSM_Tuning_Active_2009, di2024brownian} and quantitatively describes our experiments without any free parameters, as we will show in the following.

Our model treats the Cs atoms as active Brownian particles driven by inelastic spin-exchange collisions.
We describe the motion of a single Cs atom at position $\mathbf{r}_{\text{Cs}}$ via an underdamped Langevin equation in three dimensions.
The dynamics is governed by the interplay between the conservative ODT potential, the dissipative coupling arising from elastic collisions with the thermal Rb bath, and the stochastic energy injection from inelastic spin-exchange events.
The equation of motion reads

\begin{eqnarray}
    m_{\text{Cs}} \mathbf{a}_{\text{Cs}} &=& \mathbf{F}_{\text{trap}}(\mathbf{r}_{\text{Cs}}) + \mathbf{F}_{\text{drag}}(\mathbf{v}_{\text{Cs}}) \nonumber\\
    &+& \mathbf{F}_{\text{th}}(v_{\text{Cs}}) + \mathbf{F}_{\text{act}}(\tau_{\text{int}}, \mathbf{v}_{\text{Cs}}),
    \label{eq:langevin}
\end{eqnarray}
where $\mathbf{r}_{\text{Cs}}$ denotes the position vector of an individual Cs atom with acceleration $\mathbf{a}_{\text{Cs}}$ and velocity $\mathbf{v}_{\text{Cs}}$, whose magnitude is given by $v_{\text{Cs}} = |\mathbf{v}_{\text{Cs}}|$. 

The conservative force $\mathbf{F}_{\text{trap}}(\mathbf{r}_{\text{Cs}})$ models the confinement within the ODT, and results from an experimentally verified ODT model (see App.~\ref{odtmodel} for details). 
We model the effects of elastic Cs-Rb collisions through an effective frictional drag force $\mathbf{F}_{\text{drag}}(\mathbf{v}_{\text{Cs}})$ and accompanying stochastic thermal fluctuations $\mathbf{F}_{\text{th}}(v_{\text{Cs}})$. 
This effective description is well-established: the Langevin equation for a massive tracer in a dilute gas is formally derived from the microscopic kinematics of binary elastic collisions \cite{Ferrari_CP_Proper_Mobility_2000, Ferrari_CP_Particles_Dispersed_2007, Dunkel_PRE_Relativistic_Brownian_2006}, and has been successfully applied to describe the dynamics of neutral Cs impurities in a cold Rb gas \cite{Hohmann_PRL_Individual_Tracer_2017}, and even nucleon scattering in heavy-ion collisions \cite{Lin_PRC_Onebody_Langevin_2019}. 
Furthermore, the validity of this stochastic approach for cold impurities has been demonstrated in atom-ion systems, where Langevin simulations were shown to quantitatively reproduce the thermalization dynamics obtained from full event-driven molecular dynamics simulations \cite{Londono_PRA_Dynamics_Single_2022, Londono_PRA_Cold_Atomion_2023}. 
As a result of the inelastic spin-exchange collisions, the active Cs atoms can reach speeds comparable to or significantly higher than the thermal velocity of the Rb bath, so that the friction coefficient $\gamma(v_{\text{Cs}})$ is strongly velocity-dependent \cite{Hohmann_PRL_Individual_Tracer_2017, Ferrari_CP_Proper_Mobility_2000}. 
We employ a piecewise model for the drag $\mathbf{F}_{\text{drag}}(\mathbf{v}_{\text{Cs}}) = -\gamma(v_{\text{Cs}})\mathbf{v}_{\text{Cs}}$: for intermediate speeds, the friction coefficient scales quadratically with velocity ($\gamma \propto v_{\text{Cs}}^2$), while for very high speeds, it transitions to a linear dependence ($\gamma \propto v_{\text{Cs}}$).
Accordingly, the associated thermal noise $\mathbf{F}_{\text{th}}(v_{\text{Cs}})$ is modeled as Gaussian white noise with a velocity-dependent amplitude $\sqrt{2\psi^2(v_{\text{Cs}})}$ that ensures thermodynamic consistency across these regimes \cite{Dubkov_JSM_Nonlinear_Brownian_2009}; the detailed expressions for $\gamma(v_{\text{Cs}})$ and $\psi^2(v_{\text{Cs}})$ are provided in App.~\ref{SIsec:elastic_coll}.

The distinguishing feature of our system is the active force $\mathbf{F}_{\text{act}}(\tau_{\text{int}}, \mathbf{v}_{\text{Cs}})$, which we model as Poissonian shot noise \cite{Fiasconaro_JSM_Tuning_Active_2009, di2024brownian}.
This term describes the discrete momentum transfers resulting from inelastic spin-exchange collisions.
We define the active force as a sum of momentum kicks 
\begin{equation}
    \mathbf{F}_{\text{act}}(\tau_{\text{int}}, \mathbf{v}_{\text{Cs}}) = m_{\text{Cs}} \sum_{i=1}^{N_{\text{se,max}}} \Delta \mathbf{v}_{\text{se},i} \, \delta(\tau_{\text{int}} - t_i).
\end{equation}
Here, $t_i$ represents the stochastic arrival times of the spin-exchange collisions, which follow a Poisson process with a rate given by the ensemble-averaged inelastic scattering rate $\langle \overline{\Gamma_\text{se}} \rangle = \langle \overline{n_{\text{Rb}}(\mathbf{r})} \: \overline{\sigma_{\text{se}}(E_{\text{c}}) v_{\text{c}}} \rangle$.
Here, $\langle \dots \rangle$ denotes an average over all internal Zeeman states of Cs.
$N_{\text{se,max}} = 6$ is the maximum number of spin-exchange collisions for a Cs atom.
Each spin-exchange collision event imparts a velocity kick $\Delta \mathbf{v}_{\text{se},i}$ in a random direction with a magnitude $\Delta v_{\text{se}} = \sqrt{4 \mu E_\text{c} / m^2_{\text{Cs}} + 2 \mu Q / m^2_{\text{Cs}}}$.
Here, the first contribution corresponds to the energy change of a Cs atom due to an elastic collision with a Rb atom, while the second contribution comes from the energy gain from the activation energy $Q$ released during the internal state transition.
Further details on the stochastic Langevin simulations are summarized in App.~\ref{langevin}.

\section{\label{sec3} Results}
\begin{figure*}[t!] 
\centering 
\includegraphics[width=\textwidth]{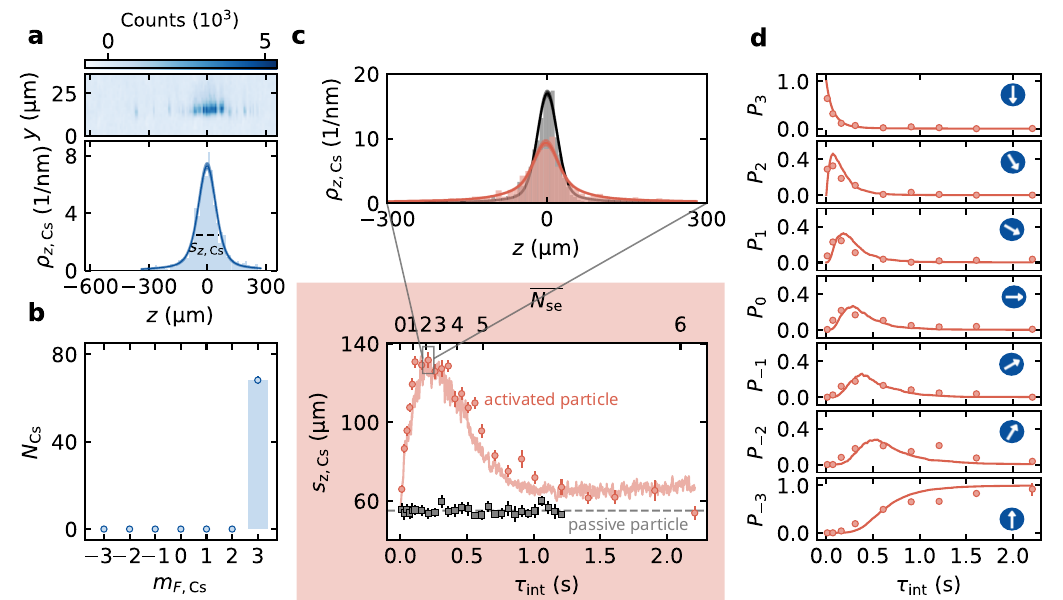} 
\caption{\textbf{Making Cs atoms active by internal state transitions.} \textbf{a} Single-shot fluorescence image of Cs atoms after $1 \, \text{s}$ in the ODT. The histogram results from 30 independent experimental runs, and the solid blue line shows a fit of the density distribution with $s_{z, \text{Cs}} = 158(2) \, \mathrm{\text{\textmu} m}$ (black dashed line) and $T_{\text{Cs}} = 2.4(1) \, \mathrm{\text{\textmu} K}$. \textbf{b} Initial Cs spin distribution before the interaction with the Rb bath. \textbf{c} Evolution of the Cs density distribution at $B = 2 \, \text{G}$. Exothermal spin-exchange with Rb atoms in $\ket{1, 0}$ activates the Cs atoms, causing a significant broadening (red dots), which is absent for passive Cs atoms in a Rb bath in $\ket{1, 1}$, where only elastic inter-species collisions are allowed (gray squares). The solid red line shows the result of an event-driven Monte-Carlo simulation. \textbf{d} Population evolution $P_{m_{f,\text{Cs}}}(\tau_{\text{int}})$ of the seven Cs spin states ($m_{f,\text{Cs}} = -3, \dots, +3$) driven by spin-exchange collisions; the solid red lines show the event-driven Monte-Carlo simulation. Technical limitations restrict the minimal interaction time to $\text{min} (\tau_{\text{int}}) = 0.01 \, \text{s}$, so the first data points do not show the initial spin-polarization in $\ket{3, 3}$ depicted in \textbf{b}. Error bars either indicate the statistical uncertainties in the atom number determination (\textbf{b} and \textbf{d}), or result from the fitting uncertainty of each density distribution fit (\textbf{c}).}
\label{fig:rb_cs_dens_evo_SE_B_var} 
\end{figure*}
In the experiments, the effect of activity on the Cs density distribution is investigated by tracking the width $s_{z,\text{Cs}}$ of the Cs ensemble in the $z$-direction for different interaction times $\tau_{\text{int}}$ with the Rb cloud (Fig.~\ref{fig:rb_cs_dens_evo_SE_B_var}).
The Cs ensemble in the (anharmonic) ODT obeys classical statistics so that its density distribution is determined by the trapping potential $U(\mathbf{r})$ and is given by $\rho_{\text{Cs}}(\mathbf{r}, T_{\text{Cs}}) = \rho_0 \exp \left(- U(\mathbf{r})/[k_{\text{B}} T_{\text{Cs}}] \right)$ with peak density $\rho_0$.
The density distribution in Fig.~\ref{fig:rb_cs_dens_evo_SE_B_var}a decays more slowly than a Gaussian distribution.
The resulting heavier tails become even more pronounced in the presence of inelastic spin-exchange collisions, since these collisions increase the kinetic energy of the Cs atoms and, consequently, their displacement from the center of the ODT.
We extract the evolution of the widths $s_{z,\text{Cs}}$ during the active diffusion by fitting the data histograms for the $z$-dependent density $\rho_{z, \text{Cs}}$ with a bimodal density distribution
\begin{align}
	\rho_{z, \text{Cs}}(z, \tau_{\text{int}}) &= \left[ 1 - w(\tau_{\text{int}})\right] \rho_{z, \text{Cs}}(z, \widetilde{T}_{\text{Cs}}) \notag\\&\quad\,+ w(\tau_{\text{int}}) \rho_{z, \text{Cs}}(z, T_{\text{Rb}}),
    \label{eq:fitfunc-cs-dens}
\end{align} 
where $w(\tau_{\text{int}})$ is a time-dependent mixing weight between 0 and 1.
The weight $w(\tau_{\text{int}})$ and the effective Cs temperature $\widetilde{T}_{\text{Cs}}$ are the only free parameters of the fit.
We choose the percentile-based width $s_{z,\text{Cs}} = z_{84} - z_{16}$ as a measure of the spread of the density distribution to include also the heavy tails.
Here, $z_p$ denotes the position of the $p$-th percentile. 
The width $s_{z,\text{Cs}}$ describes a range around $z=0$ that covers 68\% of the Cs density (analogously to the one-sigma range of a Gaussian distribution).
\begin{figure*}[t!] 
\centering 
\includegraphics[width=\textwidth]{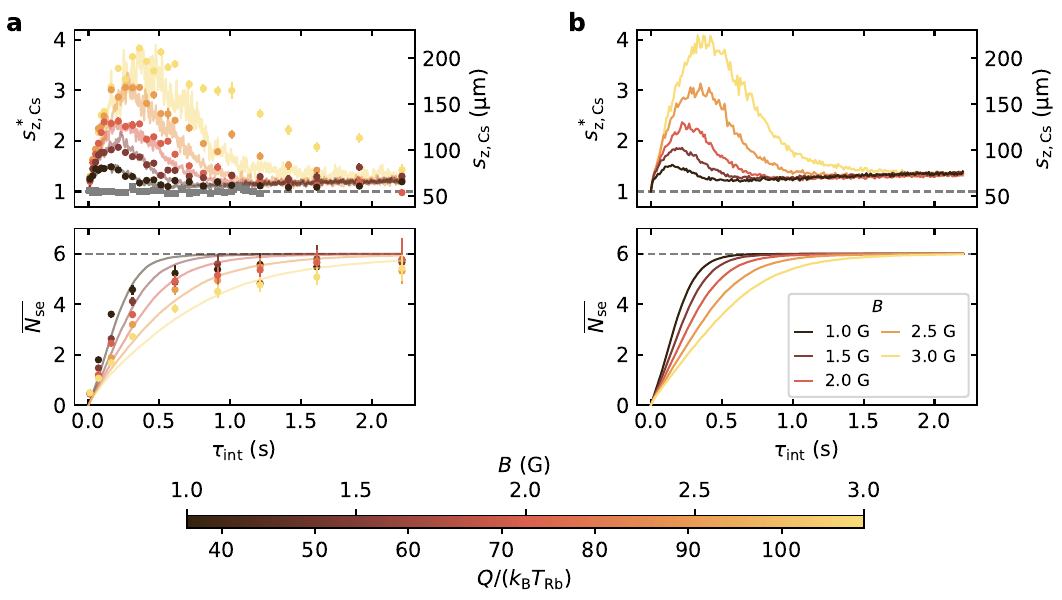} 
\caption{\textbf{Controlling the activation energy.} \textbf{a} Evolution of the measured Cs density distribution for active (dots) and passive (gray squares) atoms (top), and the corresponding number-of-collisions evolution derived from the populations of the seven Cs Zeeman states (bottom). Widths $s_{z,\text{Cs}}^*$ are normalized to the case without exothermal spin-exchange collisions (gray dashed line). Data points show averages over $\approx$ 90 independent experimental runs; solid lines correspond to event-driven Monte-Carlo collision simulations. Error bars either result from the fitting uncertainty of each density distribution fit or indicate the statistical uncertainties in the atom number. \textbf{b} Predicted Cs density distribution evolution (top) and corresponding number-of-collisions evolution $\overline{N_\mathrm{se}}(\tau_\mathrm{int})$ (bottom) for active Cs atoms from the coarse-grained active Langevin model.}
\label{fig:rb_cs_dens_evo} 
\end{figure*}

The activity of the diffusive Cs atoms results in a significant transient broadening of the density distribution, which is completely absent in a passive Cs ensemble. Figure~\ref{fig:rb_cs_dens_evo_SE_B_var}c shows an exemplary evolution of the density distribution at a fixed magnetic field of $B = 2 \, \text{G}$. 
The width $s_{z,\text{Cs}}$ first steeply increases to reach a peak value of $s_{z,\text{Cs}} \approx 130 \, \mathrm{\text{\textmu} m}$, before eventually relaxing to its initial value $s_{z,\text{Cs}} \approx 55 \, \mathrm{\text{\textmu} m}$.
This relaxation is a result of the limited number of exothermal spin-exchange collisions: a Cs atom loses its activity as soon as it reaches its uppermost Zeeman state after $N_{\text{se}} = 6$ collisions, and thus subsequently interacts only via elastic collisions with the surrounding Rb cloud.
In principle, this can be overcome by external fields bringing back the Cs population to the initial Zeeman state, i.e. refueling the system to prolong the activation period.
Consequently, the Cs ensemble thermalizes for long interaction times $\tau_{\text{int}}$ so that its spatial distribution eventually coincides with its initial distribution again.
An independent measurement of the spin evolution allows us to link the time evolution of the Cs atoms' density distribution with the quantum spin dynamics and the associated activation energy (Fig.~\ref{fig:rb_cs_dens_evo_SE_B_var}d).
The unidirectional spin-exchange drives the Cs atoms from the ground state $\ket{3, 3}$ to the uppermost Zeeman state $\ket{3, -3}$ within $\tau_{\text{int}} = 2.21 \, \text{s}$, passing through each of the seven Zeeman states. 
We find that the width $s_{z,\text{Cs}}$ of the active Cs ensemble peaks approximately after a mean number of $\overline{N_{\text{se}}} \approx 2$ spin-exchange collisions. 

\subsection*{Influence of the activation energy on the dynamics}
We now turn to a systematic investigation of the active motion and first characterize the influence of the activation energy $Q$ by fixing the temperature of the Rb cloud to $T_{\text{Rb}} = 463(10) \, \text{nK}$ and varying the magnetic field $B$ from $1 \, \text{G}$ to $3 \, \text{G}$ (Fig.~\ref{fig:rb_cs_dens_evo}).
The resulting activation energies $Q$ range from $k_{\text{B}} \times 17 \, \mathrm{\text{\textmu} K}$ to $k_{\text{B}} \times 51 \, \mathrm{\text{\textmu} K}$ [Eq.~\eqref{eq:Q}], and are comparable to the ODT depth $\abs{U_{0,\text{Cs}}} = k_{\text{B}} \times 43 \, \mathrm{\text{\textmu} K}$ of the Cs atoms. 
However, the energy $Q$ is released to the relative kinetic energy of the colliding atom pair, i.e., only the proportion $\mu / m_{\text{Cs}} \approx 0.4$ is transferred to the Cs atom.
As a result, a single spin-exchange collision does not lead to an immediate loss of the Cs atom from the ODT.
\begin{figure*}[t!] 
\centering 
\includegraphics[width=\textwidth]{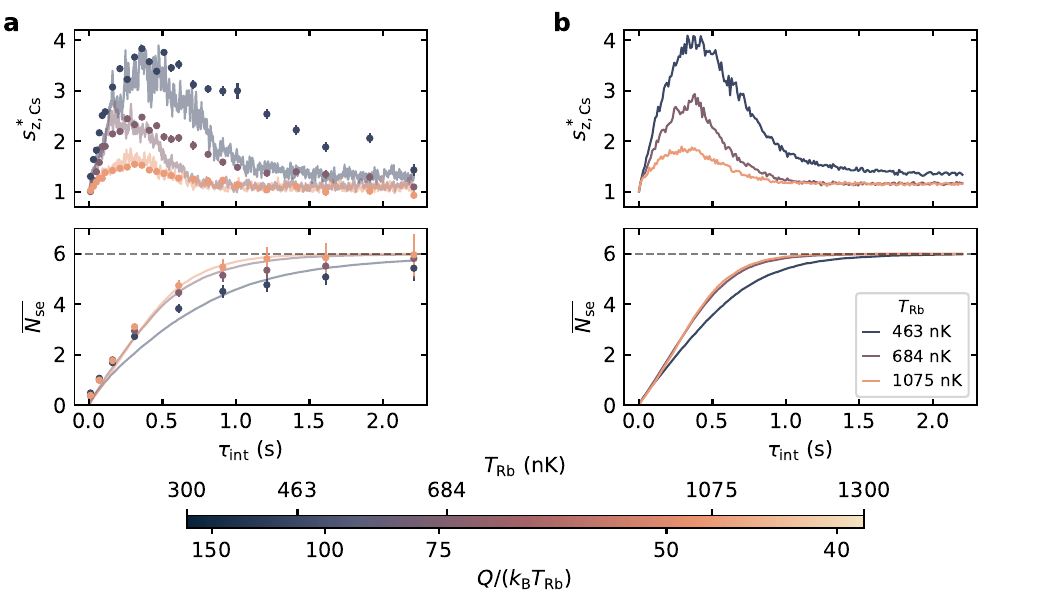} 
\caption{\textbf{Active diffusion in different temperature baths.} \textbf{a} Measured evolution of the Cs density distribution (top) and corresponding number-of-collisions evolution (bottom). Widths $s_{z,\text{Cs}}^*$ are normalized to the width of a passive Cs ensemble. Data points average $\approx$ 90 independent experimental runs; solid lines show results of event-driven Monte-Carlo simulations. Error bars either result from the fitting uncertainty of each density distribution fit or indicate the statistical uncertainties in the atom number. \textbf{b} Predicted Cs density distribution evolution (top) and corresponding number-of-collisions evolution (bottom) for active Cs atoms from the coarse-grained active Langevin model.}
\label{fig:rb_cs_dens_evo_T_var} 
\end{figure*}

Figure~\ref{fig:rb_cs_dens_evo}a shows that increasing the magnetic field -- and thus the activation energy $Q$ -- leads to a progressively stronger modification of the Cs density distribution. 
We find that the width $s_{z,\text{Cs}}$ of an active Cs ensemble can exceed that of a passive one by nearly a factor of four at the largest magnetic field applied, $B = 3 \, \text{G}$, corresponding to an activation energy of $Q = k_{\text{B}} \times 110 \, T_{\text{Rb}}$. 
This indicates that active Cs atoms can leave the Rb cloud as a result of the high energy gain in an inelastic spin-exchange collision, propagate within the ODT, and eventually re-enter the Rb cloud.

Importantly, not only does the peak width of the Cs density distribution increase with the magnetic field $B$, but the time at which the width peaks is also shifted to later times.
This delay arises from a slowed collisional interaction caused by three effects: (i) the reduced density overlap between the two atomic species, (ii) the fact that high-energy collisions happen less frequently, and (iii) the decrease of inelastic scattering cross sections with increasing $B$.

We compare our experimental data to the predictions of the event-driven Monte-Carlo collision simulations (Fig.~\ref{fig:rb_cs_dens_evo}a) and the active Langevin model (Fig.~\ref{fig:rb_cs_dens_evo}b), finding overall excellent quantitative agreement without free parameters. 
Agreement is slightly reduced for magnetic fields larger than $B = 2 \, \text{G}$.
The systematic deviations observed at larger magnetic fields $B$ can be attributed to an additional anharmonicity of the ODT that is not included in the ODT potential used in the simulations.
This enhanced anharmonicity becomes relevant only at higher $B$, where the increased activation energy $Q$ leads to larger atomic displacements from the trap center, thereby enhancing the influence of anharmonic contributions.

\subsection*{Influence of the bath temperature on the activation}
To ensure comparability across all these measurements, the mean density of the Rb cloud is held constant at $\overline{n_{\text{Rb}}} = 0.28(2) \times 10^{13} \, \text{cm}^{-3}$.
The corresponding mean collision energies $\overline{E_{\text{c}}} = 3 k_{\text{B}} T_{\text{Rb}}/2$ range from $k_{\text{B}} \times 0.6 \, \mathrm{\text{\textmu} K}$ to $k_{\text{B}} \times 1.8 \, \mathrm{\text{\textmu} K}$, which are factors of 85 to 28 smaller than $Q$.
Moreover, we investigate the influence of the Rb cloud temperature on the active motion of the Cs atoms by fixing the magnetic field at $B = 3 \, \text{G}$, corresponding to an activation energy $Q = k_{\text{B}} \times 51 \, \mathrm{\text{\textmu} K}$, and varying $T_{\text{Rb}}$ between $463(10) \, \text{nK}$ and $1075(10) \, \text{nK}$ (Fig.~\ref{fig:rb_cs_dens_evo_T_var}).

Figure~\ref{fig:rb_cs_dens_evo_T_var} shows that the modification of the Cs density distribution becomes weaker at higher Rb cloud temperatures.
This is because the collision energies are comparably high at higher temperatures, reducing the relative impact of the activation energy $Q$ on the active motion.
For instance, the width $s_{z,\text{Cs}}$ of active Cs atoms in a Rb cloud with $T_{\text{Rb}} = 1075(10) \, \text{nK}$ exceeds that of passive Cs atoms only by a factor of 1.5.

Interestingly, we observe that the spin-exchange dynamics accelerates with increasing temperature $T_{\text{Rb}}$, although the inelastic scattering cross sections suggest a slowdown.
One reason is that at low temperatures $T_{\text{Rb}}$ the Cs density distribution is strongly modified, reducing its overlap with the Rb cloud and thus slowing spin-exchange. 
Another reason is a technical heating of the Rb ensemble in the ODT, which is accompanied by Rb atom losses (see App.~\ref{lifetime}). 
This effect influences measurements with low Rb temperatures particularly strongly. 
As a result, the spin dynamics slows steadily as the interaction time $\tau_{\text{int}}$ increases.

\section{\label{sec4} Conclusion}
In conclusion, we realize individual Cs atoms as active, atom-sized diffusive particles, extracting energy from quantum spin degrees of freedom and converting it into active motion.
This mechanism is intrinsically quantum, and the activation is additionally significantly boosted by the high-energy tail of the low-energy scattering cross sections in matter wave collisions.
Our experimental observation is quantitatively modeled by a coarse-grained active Langevin model derived from kinetic theory.
The resulting dynamics so far is effectively classical, albeit at the edge of the quantum domain.
Further cooling can increase the thermal de Broglie wavelength to enhance matter wave effects, where activation and localization in time-varying potentials will compete \cite{Barbosa25, Finelli26}.
The activation is transient because the Cs spin spectrum is bound in the present case.
However, external radio-frequency radiation may be used to coherently transfer the Cs spin population back to the initial state to prolong the activation period.
Technical heating and atom loss in the Rb cloud so far impact the results on long times, but do not alter the qualitative interpretation of the quantum-enabled active dynamics.

Our work opens the route to probe the contribution of quantum effects for atom-sized active particles.
It will be interesting to probe entropy production as well as active transport in externally engineered potentials.
Moreover, by increasing the number of Cs atoms, collective phenomena of active matter close to or within the quantum domain will come into reach.
Furthermore, bringing the Rb gas to a quantum degenerate state may enable studies of anomalous diffusion and long-lived transport correlations in quantum-enabled active matter, analogous to recent studies in passive ultracold systems \cite{Sagi2012AnomalousDiffusion,Afek2017PowerLawDynamics}.
Likewise, reducing the number of Rb atoms in the bath will pave the way to realizing non-Markovian baths. 
Finally, as the quantum regime is entered with a large number of Cs atoms, it will be interesting to explore the possibility of realizing active quantum many-body systems.
An intriguing direction might be to employ the mechanism presented here to engineer a device that combines the advantages of quantum reservoir computing \cite{paparelle2026experimental} and active-matter reservoir computing \cite{teVrugt_book26, Gaimann25}.

\begin{acknowledgments}
S.B.\ acknowledges funding by the Studienstiftung des deutschen Volkes. M.t.V., B.L., R.W., and A.W.\ are funded by the Deutsche Forschungsgemeinschaft (DFG, German Research Foundation) -- Project-IDs 464588647 -- SFB 1551 (M.t.V.); 233630050 -- TRR 146 (B.L.); 535275785 (R.W.); 277625399 -- SFB/TR185 (A.W.). M.t.V. also received funding from the Carl Zeiss Foundation. Moreover, A.W. acknowledges funding by the European Research Council (ERC) under the European Union’s Horizon Europe research and innovation programme (ERC Advanced Grant No. 101200776).
\end{acknowledgments}

\section*{Author Contributions}
S.B.\ took and analyzed the experimental data and performed the numerical event-driven Monte-Carlo simulations for benchmarking the experimental data.
J.F.\ and S.H.\ helped run the experimental apparatus.
A.G.\ provided the scattering cross sections for the event-driven Monte-Carlo collision simulations. A.K.M.\ performed the Langevin dynamics simulations. S.L.\ derived the Langevin model from kinetic theory.
M.t.V., R.W., and A.W.\ conceived the idea. 
A.K.M., S.L., M.t.V., B.L., H.L., and R.W. developed the supporting theoretical model. M.t.V., B.L., H.L., R.W., and A.W. supervised the work. All authors contributed to the interpretation of data, discussion, scientific presentation, and manuscript writing. 

\section*{Data Availability}
The data that support the findings of this study are openly available at the following URL/DOI: \url{https://doi.org/10.5281/zenodo.20814314}.

\clearpage

\onecolumngrid
\appendix

\section{\label{exp-system}Experimental system}
\subsection{\label{exp-setup}Experimental sequence and observables}
The initial preparation of the two atomic samples is carried out sequentially in spatially separated regions.
In the first step, we perform an all-optical preparation of the thermal Rb cloud.
This stage consists of a laser-cooling step followed by evaporative cooling in an anisotropic, crossed optical dipole trap (ODT) formed by the intersection of two $ 1064 \, \text{nm}$ laser beams. 
The Rb sample is additionally transferred to the magnetic-insensitive state $\ket{1, 0}$ of the $5S_{1/2}$ electronic ground state during evaporation through a combination of optical pumping and a subsequent radio-frequency Landau-Zener sweep.
The final atom number $N_{\text{Rb}}$ and temperature $T_{\text{Rb}}$ of the thermal Rb cloud is inferred from standard absorption-imaging after $7 \, \text{ms}$ of time-of-flight.
The limited resolution of our imaging system prohibits direct access to the in-situ density distribution of the Rb cloud in the ODT.
Instead, the in-situ density distribution is calculated from the measured observables $N_{\text{Rb}}$ and $T_{\text{Rb}}$ with the help of our ODT model (see App.~\ref{odtmodel}).

In a second step, about 130 Cs atoms are captured in a high-gradient magneto-optical trap and loaded into a second crossed ODT.
This trap is located at an axial distance of around $200 \, \mathrm{\text{\textmu} m}$ from the first ODT, where the Rb atoms are trapped.
Subsequent degenerate Raman sideband cooling not only cools the atoms further to a temperature of $T_{\text{Cs}} = 2.4(1) \,  \mathrm{\text{\textmu} K}$, but also optically pumps them to the state $\ket{3, 3}$ of the $6S_{1/2}$ electronic ground state.
We remove residual atoms that are not in $\ket{3, 3}$ from the ODT by a cleaning scheme, which is based on the spin-selective readout in Ref.~\cite{Schmidt2018} and consists of a combination of microwave Landau-Zener sweeps and resonant laser pulses.
This stage eventually leaves a spin-polarized Cs ensemble consisting of about $N_{\text{Cs}} = 68(2)$ atoms in the absolute ground state $\ket{3, 3}$.

In the last step, the Cs atoms are transported by a species-selective, one-dimensional optical lattice formed by two counter-propagating $790 \, \text{nm}$ laser beams into the Rb cloud.
The lattice is then switched off during the following collisional interaction.
We transfer the Rb ensemble to its absolute ground state $\ket{1, 1}$ so that angular momentum conservation restricts the inter-species collisions to elastic $s$-wave collisions, and the activation via inelastic spin-exchange collisions is absent.
The Cs atoms thermalize for $\tau_{\text{th}} = 150 \, \text{ms}$ at a small magnetic field of $B = 50 \, \text{mG}$.
In order to initiate active diffusion, we restore the internal Rb state to $\ket{1, 0}$ after the thermalization, allowing spin-exchange to occur for a tunable interaction time $\tau_{\text{int}} = 0.01 \, \text{s}$ to $2.21 \, \text{s}$.

It is important to note that the one-dimensional optical lattice is not only used as a conveyor belt, but also for spatially resolved fluorescence imaging of the Cs atoms after the interaction with the Rb sample.
The lattice creates a repulsive potential for the Cs atoms along the $z$-direction, and thereby pins the atoms' position in this dimension.
Our imaging system provides a resolution of $1.4 \, \mathrm{\text{\textmu} m}$, so that single lattice sites cannot be resolved.
We can combine the spatially-resolved fluorescence imaging with a series of microwave Landau-Zener sweeps and resonant laser pulses to also gain information about the spin distribution of the Cs atoms \cite{Schmidt2018}.

\subsection{\label{odtmodel}Optical dipole trap model}
\begin{figure}[t] 
\centering 
\includegraphics[width=0.5\textwidth]{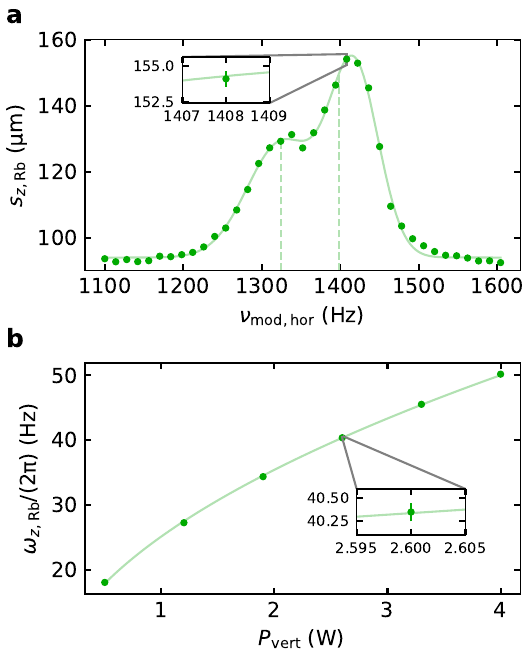} 
\caption{\textbf{Rb trapping frequencies.} \textbf{a} Parametric excitation of the Rb atoms by periodic modulation of the power $P_{\text{hor}}$ of the horizontal ODT beam with frequency $\nu_{\text{mod}, \text{hor}}$. We set the mean power of the horizontal ODT beam to $P_{\text{hor}} = 100 \, \text{mW}$ and the power of the vertical ODT beam is fixed to $P_{\text{vert}} = 1.6 \, \text{W}$. The amplitude of the power modulation of the horizontal ODT beam is set to $2.5 \, \text{mW}$ and the modulation is turned on for $\tau_{\text{mod}} = 400 \, \text{ms}$. We allow the Rb atoms to equilibrate for $\tau_{\text{eq}} = 50 \, \text{ms}$ after the modulation, and measure the size $s_{z,\text{Rb}}$ of the Rb cloud after $7 \, \text{ms}$ of time-of-flight as a function of the modulation frequency. Each data point in panel is an average of 12 independent experimental runs. The green dashed lines show the resonance positions resulting from our ODT model (Tab.~\ref{tab:ODT-params}). \textbf{b} Axial trapping frequency $\omega_{z,\text{Rb}}/(2\pi)$ of the Rb atoms for various mean powers $P_{\text{vert}}$ resulting from parametric excitation of the Rb ensemble. The power of the horizontal ODT beam is fixed to $P_{\text{hor}} = 16 \, \text{mW}$. The solid green line results from our ODT model.
The error bars either result from the fitting uncertainty of the Rb line density (\textbf{a}) or indicate the fitting uncertainty of the excitation spectrum (\textbf{b}). The insets show the typical error bar size.}
\label{fig:rb_trap_frequencies} 
\end{figure}

The ODT potential of our experiment is generated by two crossed laser beams at a wavelength of $1064 \, \text{nm}$. 
We choose a right-hand coordinate system which is oriented in such a way that gravity acts along the $x$-direction. 
One laser beam is traveling along the $z$-direction and the second one is traveling along the $x$-direction.
We employ a representation of the experimental ODT potential, which assumes Gaussian
beam shapes and includes experimentally measured beam waists. 
The vertical laser beam is modeled as a Gaussian laser beam with beam waist $w_{0,\text{vert}} = 165 \, \mathrm{\text{\textmu} m}$ and focus position $\mathbf{O}_{\text{vert}} = (52.5, 0, 0)^\mathrm{T} \, \text{mm}$. We assume the horizontal laser beam to be an elliptical Gaussian laser beam with beam waists $w_{0,\text{hor}, 1} = 20.63 \, \mathrm{\text{\textmu} m}$ and $w_{0,\text{hor}, 2} = 21.82 \, \mathrm{\text{\textmu} m}$ and focus position $\mathbf{O}_{\text{hor}} = (0, 0, 0)^\mathrm{T} \, \mathrm{\text{\textmu} m}$. 
We can adjust the shape and depth of the ODT, and thereby also the trapping frequencies of the atoms, by setting the powers $P_z = P_{\text{hor}}$ and $P_x = P_{\text{vert}}$ of the two laser beams.
To verify the model of our ODT potential, we can measure the radial and axial trapping frequencies of the Rb atoms and compare them to the trapping frequencies resulting from a harmonic approximation in our model (Fig.~\ref{fig:rb_trap_frequencies}).
We parametrically excite the Rb atoms by applying a periodic modulation of the trapping potential at a frequency $\nu_{\text{mod}, j}$ that is approximately twice the trapping frequency of direction $j$ we want to measure.
This process initially increases the kinetic energy, and therefore the oscillation amplitude of the Rb atoms, along the drive direction $j$.
The anharmonicity of the ODT potential couples the three motional degrees of freedom of the atoms so that elastic collisions between the Rb atoms redistribute energy between all three spatial directions.    
Consequently, the size of the cloud increases in all directions and not only along the drive direction $j$ if we allow the Rb atoms to equilibrate after the modulation.
We detect the size $s_{z,\text{Rb}}$ of the Rb cloud after $7 \, \text{ms}$ of time-of-flight as a function of the modulation frequency $\nu_{\text{mod}, j}$ to infer the trapping frequency of direction $j$.
Figure~\ref{fig:rb_trap_frequencies}a shows, as an example, the measurement of the two radial trapping frequencies $\omega_{x,\text{Rb}}/(2\pi)$ and $\omega_{y,\text{Rb}}/(2\pi)$.
The resonance frequencies $\nu_1 = 1326(3) \, \text{Hz}$ and $\nu_2 = 1418(1) \, \text{Hz}$ are close to the expected frequencies (green dashed lines) which were calculated from the ODT parameters in Tab.~\ref{tab:ODT-params}.
\begin{table*}[tb]
\caption[Overview ODT parameters]{ODT depth $U_0$ and trapping frequencies $\omega_k$ with $k\in\{x,y,z\}$ for the two atomic species inferred from our ODT model in harmonic approximation. The power of the laser beams is fixed to $P_{\text{hor}} = 100 \, \text{mW}$ and $P_{\text{vert}} = 1.6 \, \text{W}$. The trapping frequencies $\{\omega_{x}, \omega_{y}, \omega_{z}\}_{\text{Cs}}$ of the Cs atoms are solely based on our ODT model and have not been measured independently.}
\centering
  \begin{tabular}{ c c c c c }
    \toprule
    Atomic species & $U_0/k_{\text{B}}$ ($\mathrm{\text{\textmu} K}$)& $\omega_x/(2 \mathrm{\pi})$ (Hz) & $\omega_y/(2 \mathrm{\pi})$ (Hz) & $\omega_z/(2 \mathrm{\pi})$ (Hz) \\ \midrule
    $^{133}\text{Cs}$ & $-43$ & $737$ & $699$ & $34$\\ 
    $^{87}\text{Rb}$ & $-26$ & $699$ & $663$ & $33$\\
    \bottomrule
  \end{tabular}
  \label{tab:ODT-params}
\end{table*}
A measurement of the axial trapping frequency $\omega_{z,\text{Rb}}/(2\pi)$ for various powers $P_{\text{vert}}$ of the vertical ODT beam shows also excellent agreement with the prediction from our ODT model (Fig.~\ref{fig:rb_trap_frequencies}b).

\subsection{\label{scatteringproperties}Scattering properties}
In our experiment, the atoms interact via ultracold two-body collisions, with both Rb-Rb and Rb-Cs interactions being repulsive. 
Importantly, the Cs-Cs interaction is negligible due to the extremely low Cs density. 
We quantify the inter-species interaction at position $\mathbf{r}$ within the Rb cloud by the local scattering rate
\begin{equation}
    \Gamma_j(\mathbf{r}, N_{\text{Rb}}, E_{\text{c}}, B) = n_{\text{Rb}}(\mathbf{r}) \kappa_j (E_{\text{c}},B),
    \label{eq:scat-rate}
\end{equation}
where $E_{\text{c}} = \mu v_{\text{c}}^2/2$ is the collision energy of the colliding atom pair, and $j = \{ \text{el}, \text{se}\}$ denotes the type of the collision. 
The Rb density $n_{\text{Rb}}(\mathbf{r}) \propto N_{\text{Rb}}$ enters linearly and sets the overall timescale of the collisional dynamics, while the collision rate coefficient $\kappa_j (E_{\text{c}},B) = \sigma_j(E_{\text{c}}, B) v_{\text{c}}$ depends on the collision energy $E_{\text{c}}$ and the corresponding scattering cross section $\sigma_j(E_{\text{c}},B)$.
In thermal equilibrium, the collision energies follow a Maxwell-Boltzmann distribution \cite{Cannoni2014}
\begin{equation}
p(E_{\text{c}}, T_{\text{Rb}}) \mathrm{d} E_{\text{c}}\label{eq:distr-func-coll-energy} = \frac{2 \pi}{(\pi k_{\text{B}} T_{\text{Rb}})^{3/2}} \sqrt{E_{\text{c}}} \exp \left( - \frac{E_{\text{c}}}{k_{\text{B}} T_{\text{Rb}}} \right) \mathrm{d} E_{\text{c}},    
\end{equation}
whose properties are solely determined by the temperature $T_{\text{Rb}}$ of the Rb cloud. 
In our system, however, the kinetic energy gain due to inelastic spin-exchange collisions enhances the probability of collision energies in the tail of the distribution function, leading to a bimodal distribution function.

\begin{figure*}[tb] 
\centering 
\includegraphics[width=\textwidth]{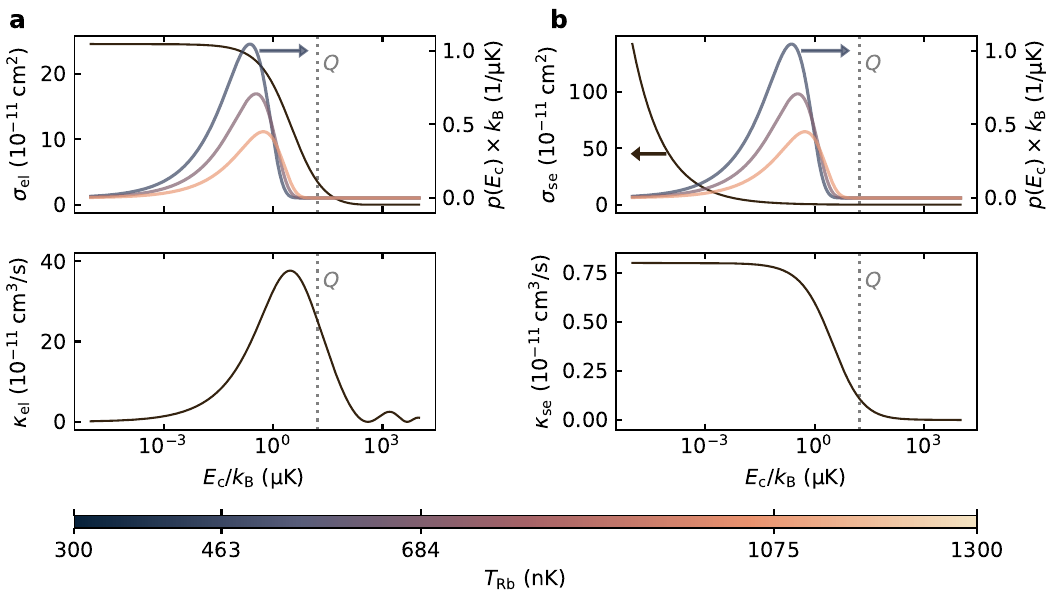} 
\caption{\textbf{Energy-dependent scattering.} Shown are the scattering cross sections $\sigma_\mathrm{el}$ and $\sigma_\mathrm{se}$ (top) and collision rate coefficients $\kappa_\mathrm{el}$ and $\kappa_\mathrm{se}$ (bottom) for \textbf{a} elastic collisions and \textbf{b} inelastic spin-exchange collisions for Cs atoms in the $\ket{3, 3}$ state with Rb atoms in the $\ket{1, 0}$ state at a fixed magnetic field $B = 1 \, \text{G}$. The upper row plots show additionally the thermal equilibrium distribution function $p(E_\mathrm{c})$ of the collision energies $E_\mathrm{c}$ for three distinct Rb cloud temperatures $T_{\text{Rb}} = \{ 463 \, \text{nK}, 684 \, \text{nK}, 1075 \, \text{nK}\}$ (refer to the right-hand axis); the transient modification of the distribution function resulting from the inelastic spin-exchange collisions is here not shown. In addition, the activation energy $Q = k_{\text{B}} \times 16.8 \, \mathrm{\text{\textmu} K}$ is also indicated (gray dotted line).}
\label{fig:rb_cs_scattering_properties_raw_mF3} 
\end{figure*}

\begin{figure*}[tb] 
\centering 
\includegraphics[width=\textwidth]{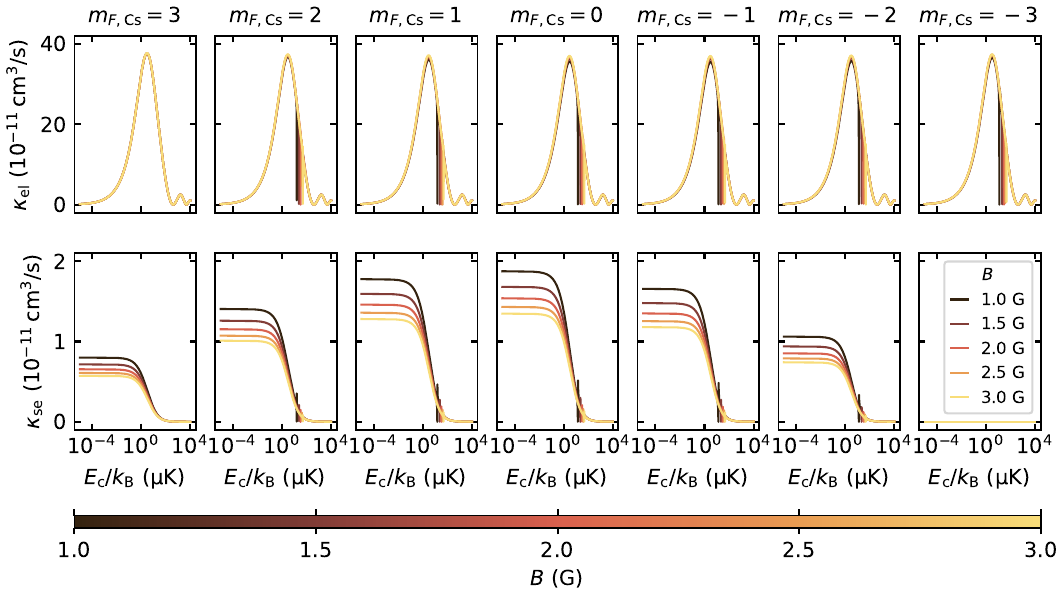} 
\caption{\textbf{Energy-dependent collision rate coefficients.} Shown are the collision rate coefficients for elastic collisions $\kappa_\mathrm{el}$ (top) and inelastic spin-exchange collisions $\kappa_\mathrm{se}$ (bottom) for Cs atoms with Rb atoms in the $\ket{1, 0}$ state as a function of the collision energy $E_\mathrm{c}$.}
\label{fig:rb_cs_scattering_properties_mF} 
\end{figure*}

The scattering cross section $\sigma_j(E_{\text{c}},B)$ is obtained from coupled-channel scattering calculations that account for the internal atomic structure of both species as well as the Rb-Cs interaction potential.
Figure~\ref{fig:rb_cs_scattering_properties_raw_mF3} shows, as an example, the energy-dependent scattering cross sections and collision rate coefficients for Cs atoms in the $\ket{3,3}$ state at $B = 1 \, \text{G}$.
The scattering is characterized by a complex energy-dependent scattering length $a(E_{\text{c}}, B) = \alpha (E_{\text{c}}, B) - i \beta (E_{\text{c}}, B)$ with magnitude $\abs{a} = \sqrt{\alpha^2 + \beta^2}$.
The resulting energy-dependent scattering cross section for elastic scattering is \cite{Hutson2007}
\begin{equation}
    \sigma_{\text{el}}(E_{\text{c}}, B) = \frac{4 \pi \abs{a}^2}{1 + k^2 \abs{a}^2 + 2 k \beta},
    \label{eq:elastic_cross_section_formula}
\end{equation}
and decreases monotonically with the collision energy $E_{\text{c}} = \hbar^2 k^2 / (2 \mu)$, where $k$ denotes the wave number associated with the relative motion of the atoms and $\hbar = h/(2\pi)$ is the reduced Planck constant.
In the zero-energy limit ($k \to 0$), the elastic scattering cross section approaches a constant value: $\sigma_{\text{el}}(E_{\text{c}}, B) \to 4 \pi \abs{a}^2$.
In the opposite limit of large $k\abs{a}$, the scattering cross section becomes independent of the scattering length: $\sigma_{\text{el}}(E_{\text{c}}, B) \to 4 \pi /k^2$.
In contrast, the inelastic scattering cross section \cite{Hutson2007}
\begin{equation}
    \sigma_{\text{se}}(E_{\text{c}}, B) = \frac{4 \pi \beta}{k \big(1 + k^2 \abs{a}^2  + 2 k \beta \big)},
    \label{eq:inelastic_cross_section_formula}
\end{equation}
diverges in the zero-energy limit ($\sigma_{\text{se}}(E_{\text{c}}, B) \propto 1/k$), while also decreasing monotonically with increasing collision energy.
It is important to note that the physically relevant quantity is the scattering rate $\Gamma_j(E_{\text{c}}, B) \propto  \kappa_j(E_{\text{c}}, B) = \sigma_j(E_{\text{c}}, B) v_{\text{c}}$, where $v_{\text{c}} = \hbar k / \mu$ denotes the collision velocity.
Consequently, the inelastic scattering rate remains finite in the zero-energy limit, despite the divergence of the corresponding scattering cross section.
Beyond their energy dependence, the collision rate coefficients $\kappa_j(E_{\text{c}}, B)$ exhibit a dependence on the external magnetic field $B$. 
Figure~\ref{fig:rb_cs_scattering_properties_mF} presents the energy-dependent collision rate coefficients for all internal Zeeman states $m_{f,\text{Cs}}$ of Cs at five distinct magnetic fields $B = \{1 \, \text{G}, \dots, 3 \, \text{G} \}$.
While the elastic collision rate coefficients show no significant variation with the applied magnetic field, the inelastic collision rate coefficients decrease in the low-energy limit as $B$ increases.  

\subsection{\label{lifetime}Rb lifetime}
\begin{figure*}[tb] 
\centering 
\includegraphics[width=\textwidth]{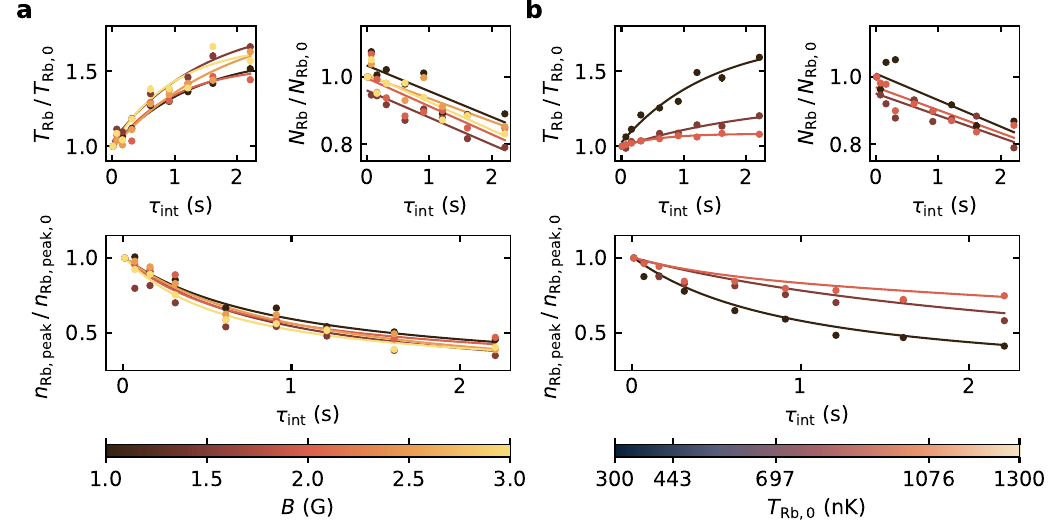} 
\caption{\textbf{Rb lifetime.} Lifetime of a Rb cloud confined in the ODT with fixed initial peak density $n_{\text{Rb}, \text{peak}, 0} \approx 0.53(2) \times 10^{13} \, \text{cm}^{-3} \approx 2 \times \overline{n_{\text{Rb},0}}$. \textbf{a} Time evolution of the characteristic parameters of a Rb cloud with initial temperature $T_{\text{Rb},0} = 463(10) \, \text{nK}$ and atom number $N_{\text{Rb},0} = 7.3(1) \times 10^3$, measured at five distinct magnetic fields $B = \{1 \, \text{G}, \dots, 3 \, \text{G} \}$ with a step size of $0.5 \, \text{G}$. \textbf{b} Time evolution of the characteristic parameters for three different initial temperatures $T_{\text{Rb},0} = \{ 443(10) \, \text{nK}, 697(10) \, \text{nK}, 1076(10) \, \text{nK}\}$ at a fixed magnetic field of $B = 3 \, \text{G}$. Solid lines represent fits according to Eqs.~\eqref{eq:fit-rb-T} and \eqref{eq:fit-rb-N}. Each data point corresponds to the average of 50 independent experimental runs. Error bars reflect the fitting uncertainty of the measured Rb line density.}
\label{fig:rb_lifetime} 
\end{figure*}

We find that the Rb cloud is constantly heated during the interaction with the Cs atoms.
The heating is accompanied by a constant loss of Rb atoms from the ODT.
Possible heating sources are recombination heating due to three-body recombination, technical noise, or spontaneous scattering of trap photons \cite{Weber2003, Gehm2000, Grimm1999}.
We detect the atom number $N_{\text{Rb}}$ and temperature $T_{\text{Rb}}$ of the Rb cloud after $7 \, \text{ms}$ of time-of-flight as a function of the interaction time $\tau_{\text{int}}$ with the Cs ensemble.
A phenomenological description of the time evolution of the Rb cloud parameters 
\begin{align}
T_{\text{Rb}}(\tau_{\text{int}})     \label{eq:fit-rb-T} &= T_{\text{Rb},\text{ss}} - (T_{\text{Rb},\text{ss}} - T_{\text{Rb},\text{0}} ) \exp \left(- \frac{\gamma_{\text{h}} \tau_{\text{int}}}{T_{\text{Rb},\text{ss}}} \right),
\\
N_{\text{Rb}}(\tau_{\text{int}}) &= N_{\text{Rb},0} - \gamma_{\text{ev}} \tau_{\text{int}}    \label{eq:fit-rb-N}
\end{align}
allows us to fit the experimental data using the free parameters $T_{\text{Rb},\text{ss}}$, $T_{\text{Rb},\text{0}}$, $\gamma_{\text{h}}$, $N_{\text{Rb},\text{0}}$ and $\gamma_{\text{ev}}$, and thus to include the heating also in our numerical simulations; the fit results are summarized in Tab.~\ref{tab:rb-lifetime}. 
Figure~\ref{fig:rb_lifetime}a shows that the heating does not depend on the chosen magnetic field $B$.
Consequently, inelastic collisions with the Cs atoms can be ruled out as a possible source of heating, since the acceleration of the atoms, and thereby also the losses, would increase with $B$.
Figure~\ref{fig:rb_lifetime}b shows, by contrast, that the heating influences cold Rb clouds particularly strongly.
\begin{table}[tb]
\caption[Fit parameters of the Rb lifetime measurements]{Fit parameters of the Rb lifetime measurements shown in Fig.~\ref{fig:rb_lifetime}.}
\centering
\begin{tabular}{c c c c c c}
\toprule
$B$ (G) & $T_{\text{Rb},0}$ (nK) & $T_{\text{Rb},\text{ss}}$ (nK) & $\gamma_0$ (nK/s) & $N_{\text{Rb},0}$ & $\gamma_{\text{ev}}$ (1/s) \\ \midrule
1.0 & 484(7) & 779(31) & 601(101) & 7368(130) & 543(122) \\
1.5 & 466(11) & 835(61) & 574(135) & 7358(95) & 613(89) \\
2.0 & 474(20) & 744(52) & 778(303) & 7363(157) & 615(147) \\
2.5 & 461(13) & 867(102) & 486(149) & 7248(115) & 489(107) \\
3.0 & 435(23) & 731(62) & 755(309) & 7214(107) & 596(100) \\ \midrule
3.0 & 455(14) & 766(74) & 534(189) & 6877(144) & 534(134) \\
3.0 & 695(8) & 914(98) & 408(243) & 13292(234) & 908(224) \\
3.0 & 1077(5) & 1166(7) & 1929(455) & 27361(421) & 1199(394) \\
\bottomrule 
\end{tabular}
\label{tab:rb-lifetime}
\end{table}

\section{Numerical simulations}
\subsection{\label{montecarlo} Event-driven Monte-Carlo collision simulations}
We use numerical event-driven Monte-Carlo collision simulations to model the active motion of individual Cs atoms inside the Rb cloud.
The Rb cloud is modeled as a thermalized, classical gas with a time-dependent temperature and density, i.e., the heating of the Rb cloud in the ODT is included, but the dynamics of the Rb cloud is not modeled explicitly.
The $s$-wave collisions between the Cs atoms and the Rb cloud are assumed to be binary and instantaneous.
Furthermore, we assume that the in-trap motion and the inter-particle collisions are decoupled over a time interval that is short compared to the mean elastic collision time, which is on the order of a few ms.
To resolve the whole system dynamics, we select a fixed simulation step width of $\delta t = 0.01/\omega_{x,\text{Cs}} \approx 2.2 \, \mathrm{\text{\textmu} s}$, which is small compared to all relevant timescales of our system.
The in-trap motion of the Cs atoms is treated classically, and the conservative potential experienced by the Cs atoms is determined by the three-dimensional ODT potential and the gravitational potential.
It is important to note that the mean-field contribution $V_{\text{MF}} = n_{\text{Rb},\text{peak}} g_{\text{Rb},\text{Cs}} \approx k_{\text{B}} \times 0.01 \, \mathrm{\text{\textmu} K}$  of the Rb cloud at maximum density $n_{\text{Rb},\text{peak}}$ to the potential of the Cs atoms is negligible compared to the typical mean collision energy $\overline{E_{\text{c}}}$, which ranges from $k_{\text{B}} \times 0.6 \, \mathrm{\text{\textmu} K}$ to $k_{\text{B}} \times 1.8 \, \mathrm{\text{\textmu} K}$, and the depth of the ODT $\abs{U_{0,\text{Cs}}} = k_{\text{B}} \times 43 \, \mathrm{\text{\textmu} K}$.
Single-particle trajectories are extracted by numerically solving Newton's equation of motion using the velocity Verlet algorithm. 
Inter-particle collisions are introduced in a Monte-Carlo manner, i.e., the probability of a collision event, as well as the energy and the direction of the particle after the collision, are determined by a sequence of random numbers.
At each discrete time step of the simulation, a random Rb collision partner, whose velocity is drawn from a Maxwell-Boltzmann distribution, is created for each Cs atom.
Subsequently, we compare the local collision probability $P_{\text{tot}} = 1 - \exp ( - \Gamma_{\text{tot}}(E_{\text{c}},B,\mathbf{r}) \delta t )$ of each Cs atom [which is determined by the local collision rate $\Gamma_{\text{tot}}(E_{\text{c}}, B,\mathbf{r})$] with a uniformly distributed random number to determine whether it will undergo a collision.
As a next step, we determine whether the collision is elastic or inelastic by creating a second uniform random number and comparing it to the fractional probability $P_j = \sigma_j/\sigma_{\text{tot}}$ of a collision of type $j = \{ \text{el}, \text{se}\}$.
The sum of these fractional probabilities is one, and consequently, the interval [0, 1] is divided
into segments of lengths corresponding to these fractional probabilities.
The type of collision is determined by the segment in which the second random number falls.
Finally, the direction of the particles after one collision is chosen randomly.
This procedure is then repeated for every time step of the simulation. 
We simulate the classical trajectories of 2500 Cs atoms simultaneously, yielding a series of the values for $\mathbf{r}$ and $\mathbf{v}$ for each atom at every discrete time step. 

\subsection{Langevin simulations}
\label{langevin}
To validate our model, we performed three-dimensional (3D) stochastic numerical simulations of Eq.~(\ref{eq:langevin}) for an ensemble of $N_{\text{Cs}} = 5000$ Cs atoms. 
We initialized the Cs atoms by sampling their positions and velocities from a Maxwell-Boltzmann distribution at initial temperature $T_{\text{Cs},0} = 2.4 \, \mathrm{\text{\textmu} K} $; the initial spatial widths $s_{j,\text{Cs}}$ ($j\in \{x,y,z\}$) in each direction were determined by the equipartition theorem, $s_{j,\text{Cs}} = \sqrt{k_\mathrm{B} T_{\text{Cs},0} / (m_{\text{Cs}} \omega_{j, \text{Cs}}^2)}$, with $m_{\text{Cs}}$ being the mass of the Cs atoms and $\omega_{j,\text{Cs}}$ their trapping frequency inferred from our ODT model (see Tab.~\ref{tab:ODT-params}).
The evolution of the Rb atom distribution is modeled using the same method as discussed in Section~\ref{montecarlo}.
We first evolved the system until a thermalization time period $\tau_{\text{th}} = 2.2 \, \text{s}$ with the magnetic field switched off, allowing the Cs ensemble to relax to equilibrium solely via elastic collisions with the Rb bath.
Following this equilibration, we enabled the spin-exchange collisions by switching on the magnetic field and simulated for a time $\tau_{\text{int}} = 2.2 \, \text{s}$.
We took the heating, and hence expansion, of the Rb bath, as well as loss of atoms during the simulation into account by implementing a time-dependent temperature of the Rb bath according to Eqs.~(\ref{eq:fit-rb-T}) and (\ref{eq:fit-rb-N}).
The values of the fit parameters in these two equations were taken from Tab.~\ref{tab:rb-lifetime}. For the simulations corresponding to $T_{\text{Rb},0}=463\, \text{nK}$ (in  Fig.~\ref{fig:rb_cs_dens_evo}b and Fig.~\ref{fig:rb_cs_dens_evo_T_var}b), we used the average values of the fit parameters across all magnetic fields (i.e., the first five rows of the Tab.~\ref{tab:rb-lifetime}): $T_{\text{Rb},\text{ss}}=791\, \text{nK}$, $\gamma_0= 639\, \text{nK/s}$ $N_{\text{Rb},0}=7310$, $\gamma_{\text{ev}}=571\, \text{s}^{-1}$. For the simulations corresponding to $T_{\text{Rb},0}=684\, \text{nK}$ and $T_{\text{Rb},0}=1075\, \text{nK}$ (in Fig.~\ref{fig:rb_cs_dens_evo_T_var}b), we used the fit parameters in the last two rows of Tab.~\ref{tab:rb-lifetime}.
We estimated the elastic and spin-exchange scattering cross-sections by fitting the velocity-dependent experimental cross-sections for each Zeeman energy level and each magnetic field using the expressions Eqs.~(\ref{eq:elastic_cross_section_formula}) and (\ref{eq:inelastic_cross_section_formula}).
The magnetic field $B$ and the initial temperature $T_{\text{Rb},0}$ of the Rb bath were varied in different simulations to match those in the experiments.
We integrated the stochastic differential equation Eq.~(\ref{eq:langevin}) using the Euler-Maruyama scheme \cite{Kim_PRE_Numerical_Method_2007}, using a timestep of $\delta t = 100 \, \mathrm{\text{\textmu} s}$. 

A list of all the parameter values used in our Langevin simulations can be found in Tab.~\ref{tab:sim_params}.
\begin{table}[!htp]
\caption{Parameters used in the stochastic Langevin simulations.}
\label{tab:sim_params}
\begin{tabular}{lll}
\toprule
\textbf{Symbol} & \textbf{Description} & \textbf{Value (units)} \\
\midrule
\multicolumn{3}{l}{\textit{Physical system \& Bath}} \\
$m_{\mathrm{Cs}}$ & Cs atom mass & $2.21 \cdot 10^{-25} \,\mathrm{kg}$ \\
$M_{\mathrm{Rb}}$ & Rb atom mass & $1.44 \cdot 10^{-25} \,\mathrm{kg}$ \\
$a$ & s-wave Rb-Cs scattering length & $3.41 \cdot 10^{-8} \,\mathrm{m}$ \\
$T_{\mathrm{Rb},0}$ & Initial Rb bath temperature & $0.463 \, \mathrm{\text{\textmu} K} - 1.075 \, \mathrm{\text{\textmu} K}$ \\
$T_{\mathrm{Cs},0}$ & Initial Cs atom temperature & $2.4 \, \mathrm{\text{\textmu} K}$ \\
$n_{\mathrm{Rb,peak,0}}$ & Initial peak Rb bath density & $5.8 \cdot 10^{18} \,\mathrm{m^{-3}}$ \\
\midrule
\multicolumn{3}{l}{\textit{Trap \& Activity}} \\
${\omega}_{\text{Rb}}/(2\pi)$ & Harmonic trap frequencies for Rb & $(699, 663, 33) \, \mathrm{Hz}$ \\
${\omega}_{\text{Cs}}/(2\pi)$ & Harmonic trap frequencies for Cs & $(737, 699, 34) \, \mathrm{Hz}$ \\
$B$ & Magnetic field & $1.0 \, \mathrm{G} - 3.0 \, \mathrm{G}$ \\
$N_{\mathrm{se,max}}$ & Maximum number of spin exchange collisions per Cs atom & $6$ \\
\midrule
\multicolumn{3}{l}{\textit{Numerical implementation}} \\
$N_{\text{Cs}}$ & Number of particles & $5000$ \\
$\delta t$ & Integration timestep & $100 \, \mathrm{\text{\textmu} s}$ \\
$\tau_{\text{th}}$ & Thermalization time without spin-exchange collisions & $2.2 \,\mathrm{s}$ \\
$\tau_{\text{int}}$ & Total simulation time with spin-exchange collisions & $2.2 \,\mathrm{s}$ \\
\bottomrule
\end{tabular}
\end{table}

\subsubsection{\label{kinetictheory}Derivation of the active Langevin model}
In this section, we derive the Langevin model given by Eq.~\eqref{eq:langevin} from a microscopic description of the gas given by kinetic theory. We make the following assumptions:
\begin{itemize}
    \item Cs-Cs interactions are negligible due to the low density of Cs.
    \item The Rb particles are in an equilibrium distribution and their magnetic moments are Kronecker-delta distributed. 
    \item Inelastic collisions with $\Delta m_F \geq 2$ can be neglected.
    \item Endothermal collisions can be neglected.
\end{itemize}
The kinetic equation considers Rb-Cs and Rb-Rb collisions. Since we want to model the Cs atom and we assume that Rb's distribution is given by an equilibrium state, we focus on the equations for Rb-Cs collisions. 
During a collision, energy and momentum are conserved.
With the velocity of the center of mass~[$\mathbf{v}_{\rm cm}\equiv (m_{\text{Cs}} \mathbf{v}_{\text{Cs}}+m_{\text{Rb}} \mathbf{v}_{\text{Rb}}) /(m_{\text{Cs}}+m_{\text{Rb}})$], energy conservation during a collision implies
{\small
\begin{align}
    &\frac{1}{2} (m_{\text{Cs}} + m_{\text{Rb}}) v_{\rm cm }^2 
    + \frac{1}{2}m_{\text{Cs}}u_{\text{Cs}}^2 +\frac{1}{2}m_{\text{Rb}} u_{\text{Rb}}^2
    + ( m_{\text{Cs}} u_{\text{Cs}} + m_{\text{Rb}} u_{\text{Rb}} )v_{\rm cm}
    + E_{\text{Cs},i} +E_{\text{Rb},j} 
    \nonumber\\
    &=\frac{1}{2} (m_{\text{Cs}} + m_{\text{Rb}}) v_{\rm cm }^2 
    + \frac{1}{2}m_{\text{Cs}} {u'}_{\text{Cs}}^2 +\frac{1}{2}m_{\text{Rb}} {u'}_{\text{Rb}}^2
    + ( m_{\text{Cs}} {u'}_{\text{Cs}} + m_{\text{Rb}} {u'}_{\text{Rb}} )v_{\rm cm}
    + E_{\text{Cs},h} +E_{\text{Rb}, k},
\label{eq:energy_cons}\end{align}}%
where $\mathbf{u}_l\equiv \mathbf{v}_l -\mathbf{v}_{\rm cm} $ is the velocity in the center-of-mass frame. 
A prime denotes the velocity after the collision.
Here, $i,h\in \{ m_{f,\text{Cs}}\}$ and $j,k\in \{ m_{f,\text{Rb}}\}$ denote the magnetic quantum numbers of Cs and Rb, respectively.
Equation \eqref{eq:energy_cons} assumes that the internal states are given by $i$ and $j$ before a collision and by $h$ and $k$ after the collision. Note that the elastic collision case can be considered by setting $h=i$ and $k=j$. 

Using momentum conservation during a collision, $m_{\text{Cs}}\mathbf{u}_{\text{Cs}} + m_{\text{Rb}}\mathbf{u}_{\text{Rb}} = m_{\text{Cs}} \mathbf{u}'_{\text{Cs}} + m_{\text{Rb}} \mathbf{u}'_{\text{Rb}}$, Eq.~\eqref{eq:energy_cons} may be written as
\begin{align}
     \frac{1}{2}m_{\text{Cs}}u_{\text{Cs}}^2 +\frac{1}{2}m_{\text{Rb}} u_{\text{Rb}}^2
    =
     \frac{1}{2}m_{\text{Cs}} {u^{\prime 2}_{\text{Cs}} } +\frac{1}{2}m_{\text{Rb}} {u^{\prime 2}_{\text{Rb}}},
    + \Delta E^{hk}_{ij} \label{eq:energy_conserv}\end{align}
where $\Delta E^{hk}_{ij}\equiv E_{\text{Cs},h}+E_{\text{Rb},k} - E_{\text{Cs},i} -E_{\text{Rb},j}$ is the change in the total internal energy during the collision. 
Since $m_{\text{Cs}} u_{\text{Cs}}^2 + m_{\text{Rb}}u_2^2= (\mu^2 /m_{\text{Cs}}) (\mathbf{v}_{\text{Cs}}- \mathbf{v}_{\text{Rb}})^2 + (\mu^2 /m_{\text{Rb}}) (\mathbf{v}_{\text{Cs}}- \mathbf{v}_{\text{Rb}})^2 = \mu (\mathbf{v}_{\text{Cs}}- \mathbf{v}_{\text{Rb}})^2$, Eq.~\eqref{eq:energy_conserv}
can be arranged in terms of the relative velocity ($\mathbf{v}_\mathrm{c} \equiv \mathbf{v}_{\text{Cs}}- \mathbf{v}_{\text{Rb}}$ )
\begin{align}
    v_{\text{c}}^2 = v^{\prime 2}_{\text{c}} + 2\Delta \frac{E^{hk}_{ij}}{\mu},
\label{eq:relative_velocity}\end{align} where $v_{\text{c}} \equiv |\mathbf v_{\text{c}}|$.
As $v_{\text{c}}$ and $v'_{\text{c}}$ are non-negative, Eq.~\eqref{eq:relative_velocity} imposes that $v^{\prime 2}_{\text{c}} + 2\Delta E^{hk}_{ij}/\mu\geq 0$ and $ {v}_{\text{c}}^2 - 2\Delta E^{hk}_{ij}/\mu\geq 0$. 

The kinetic equation~\cite{rossani1998kinetic} for the density $\rho_{i}$ of a Cs particle with internal state $i$ can be written as 
\begin{align}
    \frac{\partial \rho_{i}(\mathbf{r},\mathbf{v}_{\text{Cs}})}{\partial t} +  \nabla_x (\mathbf{v}_{\text{Cs}}\rho_{i}(\mathbf{r},\mathbf{v}_{\text{Cs}}) )+\nabla_v \bigg(\frac{f_1}{m_{\text{Cs}}}  \rho_{i}(\mathbf{r},\mathbf{v}_{\text{Cs}})\bigg) =& 
    \sum_{D_i} J^{ijhk}_i ,
\label{eq:basic_kinetic}\end{align}
where 
{\small\begin{align}
    J^{ijhk}_i (\mathbf{v}_{\text{Cs}}) =&\iint\!  \Theta( v_\mathrm{c}- \epsilon^{hk}_{ij}) v_\mathrm{c} \frac{\mathrm{d}\sigma^{hk}_{ij}}{\mathrm{d}\Omega} (v_\mathrm{c}, \Omega' ) 
    (\rho_{h} ( \mathbf{r}, \mathbf{v}^{\prime}_{\text{Cs}}) 
    \varrho_{k}( \mathbf{r}, \mathbf{v}^{\prime}_{\text{Rb} } ) 
    - \rho_{i}(\mathbf{r}, \mathbf{v}_{\text{Cs}})  \varrho_{j}(\mathbf{r}, \mathbf{v}_{\text{Rb}}) ) \,\mathrm{d}\mathbf{v}_{\text{Rb}} \mathrm{d}{\Omega'}
\label{eq:inel_collision}\end{align}}%
denotes the collision integral describing the contribution to the evolution of $\rho_{\text{Cs},i}(\mathbf{r},\mathbf{v}_{\text{Cs}})$ due to binary collisions between a Cs atom of state $i$ and Rb atom of state $j$, resulting in a final state $h$ and $k$, respectively, and $\sum_{D_i}$ denotes a summation over the set of all collision channels with a given internal state $i$. The symbols $\rho_{i}$ and $\varrho_{j}$ represent the probability density for Cs with internal state $i$ and Rb with internal state $j$. 
Equation \eqref{eq:relative_velocity} is incorporated into the collision integral $J_i^{ijhk}$ in the form of a step function $\Theta(x)$, which becomes $1$ for $x\geq 0$ and $0$ for $x<0$. 
Here, $\epsilon^{hk}_{ij}$ is the threshold that ensures that both $v_\mathrm{c}$ and $v'_\mathrm{c}$ remain positive. For $\Delta E^{hk}_{ij} > 0$, it is defined as $\epsilon^{hk}_{ij} \equiv \sqrt{2 \Delta E^{hk}_{ij}/\mu}$. For $\Delta E^{hk}_{ij}\leq 0$, we have $\epsilon^{hk}_{ij}=0$. 
$\Omega'$ denotes the solid angle for the relative velocity $\mathbf v'_c$.
Note that the double appearance of $i$ in $J_i^{ijhk}$ does not mean Einstein notation. 

Inside the integral at the right-hand side of Eq.~\eqref{eq:inel_collision}, there are $12$ variables for $\mathbf{v}_{\text{Cs}}$, $\mathbf{v}_{\text{Rb}}$, $\mathbf{v}^{\prime}_{\text{Cs}}$, and $\mathbf{v}^{\prime}_{\text{Rb}}$. 
The energy and momentum conservation laws impose four constraints. Also, $\mathbf{v}_{\text{Cs}}$ and $\mathbf{v}_{\text{Rb}}$ are given, which eliminates six undetermined variables. Finally, $\Omega$ is given as a dummy variable in the integration. Since $\Omega$ is a solid angle, it constrains two variables. Therefore, 12 variables are fully determined by the mentioned constraints ($4+6+2$). 
Equation~\eqref{eq:inel_collision} has a gain and a loss term based on microreversibility~\cite{rossani1998kinetic}. In other words, the backward collision is always possible, and the gain and loss parts are written as 
\begin{align}
     J^{ijhk}_{i,\rm gain} (\mathbf{v}_{\text{Cs}}) = \iint  \!\Theta( v'_\mathrm{c}- \epsilon^{ij}_{hk}) v'_\mathrm{c} \frac{\mathrm{d}\sigma^{ij}_{hk}}{\mathrm{d}\Omega} (v'_\mathrm{c}, \Omega) 
    \rho_h(\mathbf{r}, \mathbf{v}^{\prime}_{\text{Cs}})  \varrho_k(\mathbf{r}, \mathbf{v}^{\prime}_{\text{Rb}}) \,\mathrm{d}\mathbf{v}^{\prime}_{\text{Rb}} \mathrm{d}\Omega 
\label{eq:gain}\end{align}
and
\begin{align}
    J^{ijhk}_{i,\rm loss} (\mathbf{v}_{\text{Cs}}) = - \iint \!\Theta( v_\mathrm{c}- \epsilon^{hk}_{ij}) v_\mathrm{c} \frac{\mathrm{d}\sigma^{hk}_{ij} }{\mathrm{d}\Omega} (v_\mathrm{c}, \Omega' ) 
{\rho}_i(\mathbf{r}, \mathbf{v}_{\text{Cs}})  \varrho_j(\mathbf{r}, \mathbf{v}_{\text{Rb}}) \,\mathrm{d}\mathbf{v}_{\text{Rb}} \mathrm{d}{\Omega'}
.\label{eq:loss}\end{align}
In Eqs.~\eqref{eq:gain} and \eqref{eq:loss}, we can consider three different scenarios~\cite{rossani1998kinetic}. We first consider them for the loss term given by Eq.~\eqref{eq:loss}:

(i) Elastic collision ($h=i$ and $k=j$):
\begin{align}
    J^{ijij}_{i,\rm loss} (\mathbf{v}_{\text{Cs}}) =& -\left [ \iint \! v_\mathrm{c} \frac{\mathrm{d}\sigma^{ij}_{ij}(v_\mathrm{c} ) }{\mathrm{d}\Omega}  
 \varrho_j(\mathbf{r}, \mathbf{v}_{\text{Rb}})  \,\mathrm{d}\mathbf{v}_{\text{Rb}} \mathrm{d}{\Omega'} \right]
 {\rho}_i(\mathbf{r}, \mathbf{v}_{\text{Cs}}) 
\label{eq:collloss}.\end{align} 
In this case, the internal states are conserved, and $\sigma^{ij}_{ij}(v_\mathrm{c} )$ contributes to the elastic cross section $\sigma_{\rm el}$. 
In contrast, when the internal states change ($h\neq i$ and $k\neq j$), the corresponding processes contribute to the inelastic cross section $\sigma_{\rm se}$, arising from the spin-exchange interaction. Since we assume s-wave scattering for our ultracold gas, the differential cross section~${\mathrm{d}\sigma^{ij}_{ij} }/{\mathrm{d}\Omega}$ does not depend on the solid angle $\Omega'$, which is same for the following inelastic cases.

(ii) Exothermal collision ($h=i+1$, $k=j-1$, and $\Delta E_{ij}^{hk} < 0$): 
\begin{align}
    J^{i,j,i+1,j-1}_{i,\rm loss} (\mathbf{v}_{\text{Cs}}) =& -
    \left [ \iint  \!\Theta( v_\mathrm{c} - \epsilon^{i+1, j-1}_{ij}) v_\mathrm{c} \frac{\mathrm{d}\sigma^{i+1,j-1}_{ij} ( v_\mathrm{c} )}{\mathrm{d}\Omega}  
 \varrho_j(\mathbf{r}, \mathbf{v}_{\text{Rb}})  \,\mathrm{d}\mathbf{v}_{\text{Rb}} \mathrm{d}{\Omega'} \right]
 {\rho}_i(\mathbf{r}, \mathbf{v}_{\text{Cs}}) 
\label{eq:exo_loss_coll}.\end{align}
Here, $J^{i,j,i+1,j-1}_{i,\rm loss} $ represents the decrease in $\rho_i$ and, for exothermal collisions, the step function is always equal to $1$. 
The transition rate from the Cs state~($v_{\text{Cs}},i$) to the Cs state~($v'_{\text{Cs}}, i+1$) can be extracted using a Dirac-delta function as 
\begin{align}
    R^{i+1,j-1}_{ij}[\mathbf v'_{\text{Cs}}|\mathbf{v}_{\text{Cs}}] =&  \iint \! \Theta( v_\mathrm{c}- \epsilon^{i+1,j-1}_{ij}) v_\mathrm{c} \frac{\mathrm{d}\sigma^{i+1,j-1}_{ij} (v_\mathrm{c})}{\mathrm{d}\Omega}   \varrho_j(\mathbf{r}, \mathbf{v}_{\text{Rb}}) 
 \delta^{(3)}(\mathbf v'_{\text{Cs}} - \mathbf V'_{\text{Cs}} ( \mathbf{v}_{\text{Cs}}, \mathbf{v}_{\text{Rb}},\Omega' ) ) 
 \,\mathrm{d}\mathbf{v}_{\text{Rb}} \mathrm{d}\Omega',
\label{eq:exo}\end{align}
where $\mathbf V'_{\text{Cs}} ( \mathbf{v}_{\text{Cs}}, \mathbf{v}_{\text{Rb}},\Omega'  ) $ is the final velocity determined for the Cs atom by the energy and momentum conservation laws with given $\mathbf{v}_{\text{Cs}}$, $\mathbf{v}_{\text{Rb}}$, and $ \Omega'$. The distribution of $\mathbf v'_{\text{Cs}}$ after the collision with given $\mathbf{v}_{\text{Cs}}$ can be obtained by dividing Eq.~\eqref{eq:exo} by $W^{i+1,j-1}_{ij} (\mathbf{v}_{\text{Cs}})\equiv \int \! \mathrm{d}v'_{\text{Cs}} \, R_{ij}^{i+1,j-1}[\mathbf v'_{\text{Cs}}|\mathbf{v}_{\text{Cs}}]$:
\begin{align}
    K^{i+1,j-1}_{ij}[\mathbf v'_{\text{Cs}}|\mathbf{v}_{\text{Cs}}] \equiv \frac{R^{i+1,j-1}_{ij}[\mathbf v'_{\text{Cs}}|\mathbf{v}_{\text{Cs}}]}{W^{i+1,j-1}_{ij} (\mathbf{v}_{\text{Cs}})}.
\end{align}
In terms of $K_{ij}^{i+1,j-1}[\mathbf v'_{\text{Cs}}|\mathbf{v}_{\text{Cs}}]$ and $W^{i+1,j-1}_{ij} (\mathbf{v}_{\text{Cs}})$, Eq.~\eqref{eq:exo_loss_coll} may be expressed as 
\begin{align}
    J^{i,j,i+1,j-1}_{i,\rm loss} (\mathbf{v}_{\text{Cs}}) 
 =&-W^{i+1,j-1}_{ij} (\mathbf{v}_{\text{Cs}}) \int \!\mathrm{d}{v'_{\text{Cs}}}\, K^{i+1,j-1}_{ij}[\mathbf v'_{\text{Cs}}|\mathbf{v}_{\text{Cs}}]  \rho_i (\mathbf{r}, \mathbf{v}_{\text{Cs}})
.\end{align}

(iii) Endothermal collision ($h=i-1$, $k=j+1$, and $\Delta E_{ij}^{hk} > 0$): 
\begin{align}
    J^{i,j,i-1,j+1}_{i,\rm loss} (\mathbf{v}_{\text{Cs}}) =& -\left [ \iint  \!\Theta( v_{\text c}- \epsilon^{i-1,j+1}_{ij}) v_{\text c} \frac{\mathrm{d}\sigma^{i-1,j+1}_{ij}(v_{\text c}  )  }{\mathrm{d}\Omega} 
 \varrho_j(\mathbf{r}, \mathbf{u}_{\text{Rb}})  \,\mathrm{d}\mathbf{u}_{\text{Rb}} \mathrm{d}\Omega'\right]
 {\rho}_i(\mathbf{r}, \mathbf{u}_{\text{Cs}}) \nonumber\\
 =&-W^{i-1,j+1}_{ij} (\mathbf{v}_{\text{Cs}}) \int  \, K^{i-1,j+1}_{ij}[\mathbf v'_{\text{Cs}}|\mathbf{v}_{\text{Cs}}]  \rho_i (\mathbf{r}, \mathbf{v}_{\text{Cs}}) \,\mathrm{d} {\mathbf v'_{\text{Cs}}}
.\end{align}
Here, we assumed $s$-wave collisions, which are suited for low-energy collisions. 

Now we consider the gain terms. 
For case (i), the gain term can be rewritten as
\begin{align}
     J^{ijij}_{i,\rm gain} (\mathbf{v}_{\text{Cs}}) = \iint  \! v'_\mathrm{c} \frac{\mathrm{d}\sigma^{ij}_{ij}}{\mathrm{d}\Omega} (v'_\mathrm{c} ) 
    \rho_i(\mathbf{r}, \mathbf{v}^{\prime}_{\text{Cs}})  \varrho_j(\mathbf{r}, \mathbf{v}^{\prime}_{\text{Rb}}) \,\mathrm{d}\mathbf{v}^{\prime}_{\text{Rb}} \mathrm{d}\Omega 
\label{eq:collgain}.\end{align}

For case (ii), the gain term reads
\begin{align}
    J^{i,j,i+1,j-1}_{i,\rm gain} (\mathbf{v}_{\text{Cs}}) &= 
    \iint \! \Theta( v'_{\text c}- \epsilon^{ij}_{i+1,j-1}) v'_{\text c} \frac{\mathrm{d}\sigma^{ij}_{i+1,j-1}(v'_{\text c})}{\mathrm{d}\Omega}
    \rho_{i+1}(\mathbf{r}, \mathbf{v}^{\prime}_{\text{Cs}})  \varrho_{j-1}(\mathbf{r}, \mathbf{v}^{\prime}_{\text{Rb}})  \,\mathrm{d}\mathbf{v}^{\prime}_{\text{Rb}} \mathrm{d}\Omega  \nonumber\\
&= \int \! R^{ij}_{i+1,j-1} [v_{\text{Cs}}|v'_{\text{Cs}}] \rho_{i+1}(\mathbf{r}, \mathbf{v}^{\prime}_{\text{Cs}})  \,\mathrm{d} \mathbf{v}^{\prime}_{\text{Cs}} 
.\end{align} 
With $W^{ij}_{i+1,j-1}[\mathbf v'_{\text{Cs}}]\equiv \int \!\mathrm{d} \mathbf{v}_{\text{Cs}}\, R^{ij}_{i+1,j-1}[\mathbf{v}_{\text{Cs}}|\mathbf v'_{\text{Cs}}]$, the integral may be written as 
\begin{align}
    J^{i,j,i+1,j-1}_{i,\rm gain} (\mathbf{v}_{\text{Cs}}) 
=& \int \! W^{ij}_{i+1,j-1}[v'_{\text{Cs}}] \frac{R^{ij}_{i+1,j-1} [v_{\text{Cs}}|v'_{\text{Cs}}]}{W^{ij}_{i+1,j-1}[v'_{\text{Cs}}]} \rho_{i+1}(\mathbf{r}, \mathbf{v}^{\prime}_{\text{Cs}})  \,\mathrm{d} \mathbf{v}^{\prime}_{\text{Cs}} \nonumber\\
=& \int \! W^{ij}_{i+1,j-1}[v'_{\text{Cs}}] K^{ij}_{i+1,j-1}[ v_{\text{Cs}} | v'_{\text{Cs}}] \rho_{i+1}(\mathbf{r}, \mathbf{v}^{\prime}_{\text{Cs}})  \,\mathrm{d} \mathbf{v}^{\prime}_{\text{Cs}}. 
\end{align}
For case (iii), the gain term is rearranged as  
\begin{align}
    J^{i,j,i-1,j+1}_{i,\rm gain} (\mathbf{v}_{\text{Cs}}) 
=& \int \! W^{ij}_{i-1,j+1}[v'_{\text{Cs}}] K^{ij}_{i-1,j+1}[ v_{\text{Cs}} | v'_{\text{Cs}}] \rho_{i-1}(\mathbf{r}, \mathbf{v}^{\prime}_{\text{Cs}})  \,\mathrm{d} \mathbf{v}^{\prime}_{\text{Cs}} 
.\end{align}
Therefore, the corresponding kinetic equation is written as 
\begin{align}
    \text{[right-hand side of Eq.~\eqref{eq:basic_kinetic}]}
    =&\, J_{\rm el} -\sum_j W^{i+1,j-1}_{ij} (v_{\text{Cs}}) \int  \!K^{i+1,j-1}_{ij}[v'_{\text{Cs}}|v_{\text{Cs}}]  \rho_i (\mathbf{r}, \mathbf{v}_{\text{Cs}})  \, \mathrm{d} {\mathbf{v}'_{\text{Cs}}} \nonumber\\
    &+ \sum_j\int W^{ij}_{i+1,j-1}[v'_{\text{Cs}}] K^{ij}_{i+1,j-1}[ v_{\text{Cs}} | v'_{\text{Cs}}] \rho_{i+1}(\mathbf{r}, \mathbf{v}^{\prime}_{\text{Cs}})  \,\mathrm{d} \mathbf{v}^{\prime}_{\text{Cs}} \nonumber\\ 
    & -\sum_j W^{i-1,j+1}_{ij} (v_{\text{Cs}}) \int \!  K^{i-1,j+1}_{ij}[v'_{\text{Cs}}|v_{\text{Cs}}]  \rho_i (\mathbf{r}, \mathbf{v}_{\text{Cs}}) \, \mathrm{d} {\mathbf{v}'_{\text{Cs}}}\nonumber\\
    &+ \sum_j\int W^{ij}_{i-1,j+1}[v'_{\text{Cs}}] K^{ij}_{i-1,j+1}[ v_{\text{Cs}} | v'_{\text{Cs}}] \rho_{i-1}(\mathbf{r}, \mathbf{v}^{\prime}_{\text{Cs}})  \,\mathrm{d} \mathbf{v}^{\prime}_{\text{Cs}} 
,\end{align}
where $J_{\rm el}\equiv \sum_j [J^{ijij}_{i,\rm loss} (\mathbf{v}_{\text{Cs}})+J^{ijij}_{i,\rm gain} (\mathbf{v}_{\text{Cs}})]$ is the elastic collision integral and we assume that the elastic collision does not depend on the internal state $i$. 
Since the magnetic quantum number $m_F$ of the Rb atom is initially prepared as $j$, under the assumptions that a bath particle never undergoes a second collision once it has collided and endothermal collisions do not occur, the kinetic equation reduces to
\begin{align}
    &\frac{\partial \rho_i(\mathbf{r},\mathbf{v}_{\text{Cs}})}{\partial t} + \nabla_x (\mathbf{v}_{\text{Cs}}\rho_i(\mathbf{r},\mathbf{v}_{\text{Cs}}) )+\nabla_v \bigg(\frac{f_1}{m_{\text{Cs}}}  \rho_i(\mathbf{r},\mathbf{v}_{\text{Cs}})\bigg) \nonumber\\
    &= 
    J_{\rm el} - W^{i+1,j-1}_{ij} (\mathbf{v}_{\text{Cs}}) \int \! K^{i+1,j-1}_{ij}[\mathbf v'_{\text{Cs}}|\mathbf{v}_{\text{Cs}}]  \rho_i (\mathbf{r}, \mathbf{v}_{\text{Cs}}) \, \mathrm{d} {\mathbf{v}'_{\text{Cs}}} \nonumber\\
    &\quad\,+ \int \! W^{ij}_{i-1,j+1}(\mathbf v'_{\text{Cs}}) K^{ij}_{i-1,j+1}[\mathbf{v}_{\text{Cs}} |\mathbf v'_{\text{Cs}}] \rho_{i-1}(\mathbf{r}, \mathbf{v}^{\prime}_{\text{Cs}})  \,\mathrm{d} \mathbf{v}^{\prime}_{\text{Cs}} 
.\label{eq:fin_kinetic}\end{align}
The last two terms on the right-hand side of Eq.~\eqref{eq:fin_kinetic} represent a Markov jump process.
Elastic collisions, represented by $J_{\rm el}$, occur more frequently than inelastic collisions. Owing to the central limit theorem, the effect of elastic collisions can be approximated by a Langevin thermostat consisting of a friction term and a Gaussian noise term, following the approach of Refs.~\cite{Ferrari_CP_Proper_Mobility_2000,Ferrari_CP_Particles_Dispersed_2007}. 
We will show in App.~\ref{SIsec:elastic_coll} that the part $J_{\rm el}$ can be approximated by such
a Langevin thermostat. In other words, one can see that the kinetic equation Eq.~\eqref{eq:fin_kinetic} is the combination of a Fokker--Planck equation and Markov jump process, called differential Chapman-Kolmogorov equation. Therefore, the corresponding stochastic equation could be written as a jump-diffusion process. 
Assuming that $W_{ij}^{i+1,j-1}$ are identical for $i\in \{ -2, ... ,3 \}$, the corresponding stochastic equations are given by 
\begin{align}
     \dot {\mathbf{r}}_{\text{Cs}}(t)&= {\mathbf{v}}_{\text{Cs}}(t) , \\
    m_{\text{Cs}}\dot {\mathbf{v}}_{\text{Cs}} (t)&= -\nabla U_{\rm trap} + \mathbf{F}_{\text{drag}} + \mathbf{F}_{\text{th}}
    + \sum_\alpha \mathbf{Y}_\alpha\delta(t-t_\alpha)  \nonumber\\
    &= \mathbf{F}_{\text{trap}} + \mathbf{F}_{\text{drag}} + \mathbf{F}_{\text{th}} + \mathbf{F}_{\text{act}}
\end{align} 
with $\mathbf{F}_{\text{trap}}=-\nabla U_{\rm trap}$.
Here, $ \mathbf{F}_{\text{act}}=\sum_\alpha \mathbf{Y}_\alpha\delta(t-t_\alpha)$ corresponds to the Markov jump part in Eq.~\eqref{eq:fin_kinetic}, and $\mathbf{Y}_\alpha$ and $t_\alpha$ are sampled from $W$ and $K$. In other words, $W$ determines when the inelastic collision occurs, and $K$ determines the intensity of a Gaussian kick (impulse) induced by the inelastic collision. 
In App.~\ref{SIsec:elastic_coll}, we show the form of the elastic collision part by connecting  $\mathbf{F}_{\rm drag}$ and $\mathbf{F}_{\rm th}$ to the elastic collision cross section.

Furthermore, we approximate the size of kick~$\mathbf Y_\alpha$ and the transition rate $W$, which leads to a simplified model that still describes dynamics of our experiment. We assume that the magnitude of each kick is given by
\begin{align}
    |\mathbf Y_\alpha| = \sqrt{4 \mu {E_\text{c}} / m^2_{\text{Cs}} + 2 \mu Q / m^2_{\text{Cs}}}
\label{eq:intensity},\end{align}
as in the model of Ref.~\cite{di2024brownian}, and that the direction of the kick is uniformly random. 

As we use a high magnetic field,  $2 \mu Q / m^2_{\text{Cs}}$ term mainly contributes to the intensity of a kick. The derivation of Eq.~\eqref{eq:intensity} is as follows: Using Eq.~\eqref{eq:relative_velocity} and Eq.~\eqref{eq:Q}, one has 
\begin{align}
u_{\text{Cs}}'^{\,2} = u_{\text{Cs}}^{2} + \frac{2\mu}{m_{\rm Cs}^{2}} Q.
\label{eq:ucs}
\end{align}
Hence,
\begin{align}
|\mathbf Y_\alpha| ^{2}
= |u_{\text{Cs}}' - u_{\text{Cs}}|^{2}
= 2 u_{\rm Cs}^{2} + \frac{2\mu}{m_{\rm Cs}^{2}} Q
        - 2|u_{\text{Cs}}'|
        \left( u_{\text{Cs}}^{2} + \frac{2\mu}{m_{\rm Cs}^{2}} Q \right)^{1/2}
        \cos\theta,
\label{eq:intens1}\end{align}
where $\theta$ is the angle between $u_{\rm Cs}$ and $u'_{\rm Cs}$. Using $u_{\text{Cs}}= \frac{\mu}{m_{\rm Cs}}
\, v_c$, Eq.~\eqref{eq:intens1} is arranged as
\begin{align}
\Delta v_{\text{mag}}^{2}
=
\frac{2\mu^{2}}{m_{\rm Cs}^{2}} v_{c}^{2}
+ \frac{2\mu Q}{m^2_{\rm Cs}}- 2|u_{\text{Cs}}'|
        \left( u_{\text{Cs}}^{2} + \frac{2\mu}{m_{\rm Cs}^{2}} Q \right)^{1/2}
        \cos\theta
.\label{eq:intensity2}\end{align}
After averaging the RHS of Eq.~\eqref{eq:intensity2}, ignoring an oscillating term $- 2|u_{\text{Cs}}'|
        ( u_{\text{Cs}}^{2} + \frac{2\mu}{m_{\rm Cs}^{2}} Q )^{1/2}
        \cos\theta$ and expressing $v_c$ in terms of $E_c$, 
one can obtain Eq.~\eqref{eq:intensity}.

Also, we assume that the density of Rb atoms and the collision rate coefficient are uncorrelated, and that $W$ is independent of the system's internal state $m_F$ except for $i=-3$.
Therefore, we use an inelastic cross section $\sigma_{\text se}$ averaged over system's internal state, instead of $\sigma^{i+1,j-1}_{i,j}$. 
Finally, the transition rate can be expressed as
\begin{align}
    W \simeq  n_{\text{Rb}}(\mathbf r_{\text{Cs}})  \langle v_c \langle  \sigma_{\text se}  \rangle_{m_F}(E_c, B)  \rangle
    = \langle \overline{\Gamma_\text{se}} \rangle, 
\end{align}
where the density of the Rb gas is given by
\begin{align}
    n_{\text{Rb} }(\mathbf r)\equiv \int \! \varrho_j(\mathbf{r}, \mathbf{v}_{\text{Rb}}) \,\mathrm{d}\mathbf{v}_{\text{Rb}} 
,\end{align}
and the averaged inelastic differential cross section is given by $\langle\sigma_{\mathrm{se}}\rangle_{m_F}(E_\mathrm{c},B)$.
\subsubsection{Effective friction and noise due to elastic collisions}
\label{SIsec:elastic_coll}
The general formula \cite{Ferrari_CP_Proper_Mobility_2000, Ferrari_CP_Particles_Dispersed_2007} for deriving the friction coefficient for a test particle atom (Cs) of mass $m_{\text{Cs}}$ with velocity $\mathbf{v}_{\text{Cs}}$ moving through a cloud of gas atoms (Rb) of mass $m_{\text{Rb}}$, number density $n_{\text{Rb}}$, and velocity $\mathbf{v}_{\text{Rb}}$ is
\begin{equation}
\begin{split}
    \gamma(v_{\text{Cs}}) \mathbf{v}_{\text{Cs}} 
    &= 4 \pi^2 n_{\text{Rb}} \mathbf{v}_{\text{Cs}} \frac{m_{\text{Cs}} m_{\text{Rb}}}{m_{\text{Cs}}+m_{\text{Rb}}}\left(\frac{m_{\text{Rb}}}{2 \pi k_\mathrm{B} T_{\text{Rb}}}\right)^{3/2} \int_0^{\infty}\!\!\!\exp\!\left(-\frac{m_{\text{Rb}} v_{\text{Rb}}^2}{2 k_\mathrm{B} T_{\text{Rb}}}\right) v_{\text{Rb}}^2 \,\mathrm{d} v_{\text{Rb}} \\
    &\quad\, \int_0^\pi \!\! v_c\left(1-\frac{v_{\text{Rb}}}{v_{\text{Cs}}} \cos (\phi)\right) \sin(\phi)\, \mathrm{d} \phi 
    \int_0^\pi \!\! \sigma_{\text{el}}(v_c, \chi)[1-\cos (\chi)] \sin(\chi)\, \mathrm{d}\chi\\
    &= 4 \pi^2 \mathbf{v}_{\text{Cs}} \frac{m_{\text{Cs}} m_{\text{Rb}}}{m_{\text{Cs}}+m_{\text{Rb}}} I_{v_{\text{Rb}}} I_\phi I_\chi
\end{split}
\end{equation}
where we use the symbols $I_{v_{\text{Rb}}}, I_\phi$ and $ I_\chi$ to denote the three integrals over the variables $v_{\text{Rb}},\phi$ and $\chi$, respectively. Here, $v_c=|\mathbf{v}_{\text{Rb}}-\mathbf{v}_{\text{Cs}}|=\left(v_{\text{Rb}}^2+v_{\text{Cs}}^2-2 v_{\text{Rb}} v_{\text{Cs}} \cos \phi\right)^{1 / 2}$ being the relative velocity.
Previously, the effective friction coefficient and the form of the associated noise have been derived for the regimes where $v_{\text{Cs}} \ll v_{\text{Rb}}$ (thermal speed of bath) and $v_{\text{Cs}} \lesssim v_{\text{Rb}}$ \cite{Ferrari_CP_Proper_Mobility_2000, Ferrari_CP_Particles_Dispersed_2007, Hohmann_PRL_Individual_Tracer_2017}. These expressions are given by
\begin{equation}
\gamma_0 = n_{\text{Rb}} \sigma_{\text{el}} \frac{8}{3} \sqrt{\frac{2 m_{\text{Rb}} k_\mathrm{B} T_{\text{Rb}}}{\pi}} \frac{m_{\text{Cs}}}{m_{\text{Cs}}+m_{\text{Rb}}}, \quad \mathbf{F}_{\text{th}} = \sqrt{2 \gamma_0 k_\mathrm{B} T_{\text{Rb}}}\ \boldsymbol{\eta}(t) \quad \text{for} \quad v_{\text{Cs}} \ll v_{\text{Rb}}
\end{equation}
and
\begin{equation}
\gamma(v_{\text{Cs}})=n_{\text{Rb}} \sigma_{\text{el}} \frac{8}{15} \sqrt{\frac{2 m_{\text{Rb}}}{\pi k_\mathrm{B} T_{\text{Rb}}}} \frac{m_{\text{Cs}}\left(m_{\text{Rb}} v_{\text{Cs}}^2 / 2+5 k_\mathrm{B} T_{\text{Rb}}\right)}{m_{\text{Cs}}+m_{\text{Rb}}}, \quad \mathbf{F}_{\text{th}} = \sqrt{2\psi^2(v_{\text{Cs}})}\ \boldsymbol{\eta}(t) \quad \text{for} \quad v_{\text{Cs}} \lesssim v_{\text{Rb}}
\end{equation}
with
\begin{equation}
\begin{aligned}
\psi^2(v_{\text{Cs}}) & =2 m_{\text{Cs}} e^{\frac{m_{\text{Cs}} v_{\text{Cs}}^2}{k_\mathrm{B} T_{\text{Rb}}}} \int_{v_{\text{Cs}}}^{\infty}\!\! -\gamma(u) u e^{-\frac{m_{\text{Cs}} u^2}{k_\mathrm{B} T_{\text{Rb}}}} \,\mathrm{d}u \\
& = 2^{\frac{5}{2}} n_{\text{Rb}} \sigma_{\text{el}} \sqrt{\frac{m_{\text{Rb}} k_\mathrm{B} T_{\text{Rb}}}{\pi}} \frac{m_{\text{Rb}} m_{\text{Cs}} v_{\text{Cs}}^2+(m_{\text{Rb}}+10 m_{\text{Cs}}) k_\mathrm{B} T_{\text{Rb}}}{15(m_{\text{Rb}}+m_{\text{Cs}})},
\end{aligned}
\end{equation}
where $\boldsymbol{\eta}(t)$ denotes Gaussian white noise of zero mean and unit variance.
In the limit where the active Cs atoms are much faster than the Rb bath atoms ($v_{\text{Cs}} \gg v_{\text{Rb}}$), we use the approximation $v_c\simeq v_{\text{Rb}}[1-\frac{v_{\text{Rb}}}{v_{\text{Cs}}}\cos(\phi)]$.
The angular integral $I_\phi$ becomes
\begin{equation}
\begin{split}
    I_\phi = \int_0^\pi \!\! v_c\left(1-\frac{v_{\text{Rb}}}{v_{\text{Cs}}} \cos (\phi)\right) \sin(\phi) \,\mathrm{d}\phi 
    &\simeq \int_0^\pi \!\! v_{\text{Rb}}\left(1-\frac{v_{\text{Rb}}}{v_{\text{Cs}}} \cos (\phi)\right)^2 \sin(\phi) \mathrm{d} \phi \\
    &\simeq \int_0^\pi \!\! v_{\text{Rb}}\left(1-\frac{2v_{\text{Rb}}}{v_{\text{Cs}}} \cos (\phi)\right) \sin(\phi) \,\mathrm{d}\phi =2v_{\text{Rb}}.
\end{split}
\end{equation}
Using the hard-sphere relation $\sigma_{\text{el}}(v_c, \chi) = \frac{\sigma_{\text{el}}}{4\pi}$ (where $\sigma_{\text{el}}=4\pi a^2$, with $a$ being the s-wave Rb-Cs scattering length), the scattering angle integral $I_\chi$ is
\begin{equation}
\begin{split}
    I_\chi &= \int_0^\pi \!\! \sigma_{\text{el}}(v_c, \chi)[1-\cos(\chi)] \sin(\chi)\, \mathrm{d}\chi 
    = \frac{\sigma_{\text{el}}}{4\pi} \int_0^\pi \!\! [1-\cos(\chi)] \sin(\chi)\,\mathrm{d}\chi 
    = \frac{2\sigma_{\text{el}}}{4\pi}.
\end{split}
\end{equation}
Finally, the velocity integral $I_{v_{\text{Rb}}}$ evaluates to
\begin{equation}
\begin{split}
    I_{v_{\text{Rb}}} &= n_{\text{Rb}} \left(\frac{m_{\text{Rb}}}{2 \pi k_\mathrm{B} T_{\text{Rb}}}\right)^{3/2} \int_0^{\infty} \!\!\!\! \exp\! \left(-\frac{m_{\text{Rb}} v_{\text{Rb}}^2}{2 k_\mathrm{B} T_{\text{Rb}}}\right) v_{\text{Rb}}^2 \,\mathrm{d}v_{\text{Rb}} 
    = \frac{n_{\text{Rb}}}{4\pi}.
\end{split}
\end{equation}
Combining these terms yields the high-speed friction coefficient $\gamma(v_{\text{Cs}}) = n_{\text{Rb}} \sigma_{\text{el}} \frac{m_{\text{Cs}} m_{\text{Rb}}}{m_{\text{Cs}}+m_{\text{Rb}}} v_{\text{Cs}}$. Thus, in the active regime, the friction coefficient depends linearly on the Cs atom speed $v_{\text{Cs}}$.

The associated noise amplitude is also modified due to the velocity-dependent friction and calculated using the formalism from Ref.\ \cite{Dubkov_JSM_Nonlinear_Brownian_2009}. The stochastic thermal force is given by
\begin{equation}
    \mathbf{F}_{\text{th}}(t) = \sqrt{2\psi^2(v_{\text{Cs}})}\, \boldsymbol{\eta}(t)
\end{equation}
with the noise strength $\psi^2(v_{\text{Cs}})$ in the high-velocity limit
\begin{equation}
    \psi^2(v_{\text{Cs}}) = n_{\text{Rb}} \sigma_{\text{el}} \frac{m_{\text{Cs}} m_{\text{Rb}}}{m_{\text{Cs}}+m_{\text{Rb}}} \left[ k_\mathrm{B} T_{\text{Rb}} v_{\text{Cs}} + \frac{\sqrt{\pi}}{2} \frac{(k_\mathrm{B} T_{\text{Rb}})^{3 / 2}}{\sqrt{m_{\text{Cs}}}} e^{\frac{m_{\text{Cs}} v_{\text{Cs}}^2}{k_\mathrm{B} T_{\text{Rb}}}} \operatorname{erfc}\!\left(v_{\text{Cs}} \sqrt{\frac{m_{\text{Cs}}}{k_\mathrm{B} T_{\text{Rb}}}}\right)\! \right],
\end{equation}
where $\operatorname{erfc}$ is the complementary error function and $\boldsymbol{\eta}(t)$ is a Gaussian white noise vector with zero mean and unit variance.

\clearpage

\twocolumngrid
\bibliography{bibliography}

\end{document}